\documentclass{aastex}
\usepackage{spr-astr-addons}
\usepackage{url}
\usepackage{graphicx}
\usepackage{epsfig}
\usepackage{epstopdf}
\usepackage{amsmath}

\RequirePackage{color}

\begin{document}

\title{Gamma Ray Heating and Neutrino Cooling Rates due to Weak Interaction Processes on $sd$-shell Nuclei in Stellar Cores}
\slugcomment{Not to appear in Nonlearned J., 45.}
\shorttitle{Short article title}
\shortauthors{Autors et al.}

\author{Muhammad Fayaz\altaffilmark{1}} \and \author{Jameel-Un Nabi\altaffilmark{1}}
\and \author{Muhammad Majid\altaffilmark{1}}

\email{jameel@giki.edu.pk}

\altaffiltext{1}{Faculty of Engineering Sciences,\\GIK Institute of
Engineering Sciences and Technology, Topi 23640, Khyber Pakhtunkhwa,
Pakistan.} \altaffiltext{1}{Corresponding author email :
jameel@giki.edu.pk}

\begin{abstract}
Gamma ray heating and neutrino cooling rates, due to weak
interaction processes, on $sd$-shell nuclei in stellar core are
calculated using the proton neutron quasiparticle random phase
approximation  theory. The recent extensive experimental mass
compilation of \citep{Wang12}, other improved model input parameters
including nuclear quadrupole deformation \citep{Ram01},
\citep{Mol16} and physical constants are taken into account in the
current calculation. The purpose of this work is two fold, one is to
improve the earlier calculation of weak rates performed by
\citep{Nabi99} using the same theory. We further compare our results
with previous calculations. The selected $sd$-shell nuclei,
considered in this work, are of special interest for the evolution
of O-Ne-Mg core in 8-10 M$_\odot$ stars due to competitive gamma ray
heating rates and cooling by URCA processes. The outcome of these
competitions is to determine, whether the stars end up as a white
dwarf \citep{Nabi08}, an electron-capture supernova \citep{Jones13}
or Fe core-collapse supernova \citep{Suz16}. The selected $sd$-shell
nuclei for calculation of associated weak-interaction rates include
$^{20,23}$O, $^{20,23}$F, $^{20,23,24}$Ne, $^{20,23-25}$Na, and
$^{23-25}$Mg. The cooling and heating rates are calculated for
density range ($10 \leq \rho($\;g.cm$^{-3}) \leq $ 10$^{11}$) and
temperature range ($0.01\times10^{9}$ $\leq$ $\;T(K)$ $\leq$
$30\times10^{9}$). The calculated gamma heating rates are orders of
magnitude bigger than the shell model rates (except for $^{25}$Mg at
low densities). At high temperatures the gamma heating rates are in
reasonable agreement.  The calculated cooling rates are up to an
order of magnitude bigger for odd-A nuclei.

\end{abstract}


\section{Introduction}

In stars, as massive as (8--10) M$_\odot$, the helium burning phase
brings some crucial changes for further evolution of the core
\citep{Woosley02}: 1- The carbon and oxygen (C-O) core grows via
triple alpha reaction and alpha capture on carbon synthesizes
oxygen. 2- Subsequent $\alpha$-capture causes the formation of
$^{20}$Ne and a little $^{24}$Mg. After the helium burning phase,
the temperature and density increase progressively and can evolve
into advance burning stages. The core of carbon and oxygen
contracts, the temperature reaches $\sim5\times10^8$K and density
$\sim3\times10^6$\;g.\;cm$^{-3}$\citep{Ryan2010}. The carbon-carbon
fusion begins to create neon, sodium and magnesium as
follows: \\
$^{12}C + ^{12}C \rightarrow^{20}Ne + ^{4}He$\\
$^{12}C+^{12}C \rightarrow^{23}Na + p$ \\ $^{12}C + ^{12}
C\rightarrow^{23}Mg + n$.\\
The free protons thus produced, may add onto the sodium as: \\
$^{23}Na + ^{1}H \rightarrow^{20}Ne + \alpha$ or (alternatively
\\$^{23}Na + ^{1}H\rightarrow^{24}$Mg \citep{Halabi14}).

These reactions show that the core of massive star, after carbon
burning phase, comprises mainly of oxygen-16 ($^{16}$O), $^{20}$Ne
and $^{24}$Mg. $^{16}$O, due to its high Coulomb barrier, survives
the carbon burning phase and later captures
$\alpha$-particles to transform into $^{20}$Ne . Other derivative
reactions of carbon burning phase are the neutron-capture,
proton-capture and alpha captures to synthesize isotopes like
$^{21}$Ne, $^{22}$Ne, $^{25}$Mg, $^{26}$Al (unstable), $^{29}$Si,
$^{30}$Si and $^{31}$P \citep{Woosley02}. The relative abundance of
these species can be seen in Ref. \citep{Gar97}.

$^{16}$O being a double magic nucleus, is therefore very stable and
its fusion can not trigger at temperatures below $2\times10^{9}K$.
Before the oxygen fusion phase, at temperature
$\sim1.2-1.8\times10^{9}$K, neon burning takes place; by absorbing
high energy gamma ray and disintegrates to $^{16}$O and $\alpha$
particle spontaneously\\$^{20}Ne +\gamma\rightarrow^{16}O + ^{4}He$.
\\The helium nuclei thus produced,  react with other
existing neon nuclei as \citep{Ryan2010}:\\
$^{20}Ne + ^{4}He\rightarrow^{24}Mg +\gamma$. \\At the end of neon
burning phase, the stellar core mainly comprises of $^{16}$O and
$^{24}$Mg nuclei. In high mass stars ($>10M_\odot$), the core
temperature can exceed 2$\times10^{9}K$, and oxygen burning (O-O
fusion) can trigger via the reaction:
\\$^{16}O + ^{16}O\rightarrow^{28}Si + ^{4}He$.

Stars in mass range 8-12 M$_\odot$ may not proceed towards oxygen
burning phase and are thought to form an O-Ne-Mg core interior to a
helium core of 2-3 M$_\odot$ after the carbon burning phase
\citep{Miyaji80}. In later works, e.g. \citep{W&W80},
\citep{Nomoto87} and \citep{M&N87}, the favorable mass range for an
O-Ne-Mg core shifted to 8-10 M$_\odot$. They reported that stars in
this mass range develop an electron-degenerate O-Ne-Mg core after
the carbon burning. The electron captures begin on $^{24}$Mg and
$^{24}$Na first, and later on $^{20}$Ne and $^{20}$F. This causes a
loss of degeneracy pressure support \citep{Mar14}. Such stars can
end up their lives in variety of ways, e.g., as O-Ne-Mg white dwarfs
without explosion or accretion induced collapse, an electron-capture
supernovae (ECSN) or Fe core-collapse supernovae (Fe-CCSN)
\citep{Suz16}. Supernovae Type Ia (SNe Ia) are, supposedly the
explosive events of accreting white dwarfs. Due to relatively high
Fermi energy of the degenerate electron gas, such white dwarfs
trigger the electron capture process and  reduce the lepton fraction
significantly \citep{Nabi08}.

In ECSN or Fe-CCSN scenario, the electron capture causes a rapid
contraction of the electron-degenerate O-Ne-Mg core. The competing
processes like contraction; cooling, and heating control the
evolutionary changes in the central density and temperature,
\citep{Suz16}. If heating is fast enough relative to contraction,
the temperature would become high enough to ignite Ne-O burning,
which would proceed towards the Fe core formation. If cooling is
fast, the contraction would lead to the collapse of the O-Ne-Mg core
and an electron-capture supernova. The last episode of the life of
8-10 M$_\odot$ stars is thus set by these competitions, which are
induced by electron capture and $\beta$-decay. Therefore the
accurate weak rates and, related neutrino cooling and gamma ray
heating rates are of decisive importance for determination of the
thermodynamic and hydrodynamic trajectory of stars in the late
stages of their evolution. Such stars, ultimately, ignite Ne and O
explosively at central density $\sim2\times10^{10}$ g.cm$^{-3}$ and
the core approaches to Chandrasekhar mass \citep{Gut05} and evolve
into core collapse scenario due to electron capture on $^{20}$Ne and
$^{24}$Mg \citep{Nomoto87}.

According to Ref. \citep{Miyaji80}, much entropy is produced by
$\gamma$ ray emission from the excited states of daughter nuclei and
by distortion in electron distribution function, as the after effect
of electron capture. The emitted $\gamma$ rays heat up the core
material and cause the convection as well \citep{Takahara89}. The
temperature of the core is now high enough to ignite oxygen
deflagration. The O-Ne-Mg core grows towards the Chandrasekhar mass,
the electron capture on the post-deflagration material continues,
which determines either explosion or collapse, as the final outcome
of the star.

Weak interaction processes in massive stars constitute the energy
budget in the last stages of stellar life. Ref. \citep{Gupta07} have
shown that cooling by weak interaction neutrinos is  remarkably
reduced by the heating effect due to electron captures and the
following $\gamma$-ray emission. Owing to core mass and its
composition, these reactions can release up to a factor of 10 times
more heat at densities $\sim 10^{11}$ g.cm$^{-3}$. Thus neutrinos
take enormous energy from the transparent core of the star for
stellar densities $\sim 10^{11}$ g.cm$^{-3}$ but gamma rays, on the
other hand, are trapped in the stellar core and heat up the matter.
This competition plays its part towards the oxygen deflagration and
initiate convection in the core as well.

In this work gamma heating and (anti)neutrino cooling rates, due to
weak rates on $sd$-shell nuclei, are calculated using proton-neutron
Quasipaticle Random Phase Approximation (pn-QRPA). Extensive
calculation of stellar weak interaction rates and associated gamma
heating rates for 709 nuclei with A = 18 to 100, taking $sd$-shell
nuclei with A = 18--39 and Z = 10--20 as first phase of the project,
using the pn-QRPA theory was carried out by Nabi and Klapdor
~\citep{Nabi99,Nabi04}. In addition, these calculations also
included neutrino cooling rates, $\beta$-delayed proton(neutron)
emissions probabilities and energy rates of proton(neutron)
emissions. In later works, these rates were calculated by using
latest experimental data, fine tuning of the model parameters and
refined algorithms ~\citep{Nabi99a}-\citep{Nabi08}. In subsequent
papers (see \citep{Nabi07}, \citep{Nabi08a} and \citep{Nabi08}) the
authors calculated the Gamow-Teller transitions, the electron
capture rates and neutrino cooling rates for $^{24}$Mg.

In the current work, for the sake of direct comparison, we use the
same density and temperature range and grid points as found in
earlier works on heating and cooling rates for $sd$-shell nuclei
\citep{Oda94}  and \citep{Takahara89} who used a range as narrow as
(9.00 $\leq$ $\log \rho \text{Y}_{e}$(\;g\;cm$^{-3}$) $\leq$11.00)
and (8.20 $\leq$ $\log\;T(K)$ $\leq$ $9.60$). These authors have
calculated these rates for $sd$-shell nuclei in full shell model
configuration using effective interaction of Wildenthal
\citep{Wil84}. Other computational works on electron capture and
$\beta$-decay rates in this connection by \citep{Mar14} using large
scale shell model (LSSM) and by \citep{Suz16}, using the shell-model
calculation in $sd$-shell with USDB Hamiltonian are also worth
mentioning. The neutrino cooling rates in URCA pair with A=23 and 25
are crucial to determine the cooling rate of O-Ne-Mg core. In
contrast, nuclei with A=20 and 24 are the key species for the
core-contraction and heat generation rates in the core
\citep{Suz16}.

Our present work is devoted to a detailed analysis of the neutrino
cooling  and gamma heating rates due to weak-interaction reactions
on $sd$-shell nuclei in stellar environment. The current work
employs recent values of physical constants, mass data and model
input parameters as compared to the earlier work of \citep{Nabi99}.
In addition we calculate neutrino cooling  and gamma heating rates
for nuclei with Z = 8 and 9 (not calculated in \citep{Nabi99}). The
calculated rates can be associated with four different
weak-interaction processes i.e. the $\beta^\pm$-decay and electron
and positron captures. It is to be noted that there can be a fifth
mode in which all parent excited states decay to ground state
through gamma transition when $E_{i}\leq S_{n}$ or $S_{p}$ (not
included in this work).

In Section 2, we present the briefs of formalism for electron
capture and $\beta$-decay rates as well as neutrino cooling and
$\gamma$-ray heating rates. In Section 3, we discuss the results and
compare them with previous works. These include the earlier pn-QRPA
calculation \citep{Nabi99} and shell model calculations
\citep{Oda94} and \citep{Suz16}. The summary and conclusions are
given in Section 4.
\section{Model Description and Formalism}

In this section, a brief description of the pn-QRPA model is
presented. To save space the detailed model formalism is not
reproduced here, for which we refer to \citep{Nabi99}.
\subsection{Hamiltonian and model parameters:}
The Hamiltonian of the pn-QRPA model is given by
\begin{equation}
H^{QRPA} = H^{sp} + V^{pair} + V ^{ph}_{GT} + V^{pp}_{GT}.
\label{Eqt. 9}
\end{equation}
Wave functions and single particle energies were calculated using
the Nilsson model (axially deformed). Pairing in nuclei was treated
within the BCS approximation. The proton-neutron residual
interaction occurs through particle-particle ($pp$) and
particle-hole ($ph$) channels. In pn-QRPA, these interaction (force)
terms were given a separable form as  $V_{GT}^{ph}$ for the
particle-hole Gamow-Teller force denoted by model parameter $\chi$
and $V_{GT}^{pp}$ for the particle-particle Gamow-Teller force
denoted by model parameter $\kappa$. In this work, the values of
$\chi$ and $\kappa$ were optimized to produce terrestrial half lives
with a percent deviation range of about 10$\%$. These values of
$\chi$ and $\kappa$ resulted in fulfillment of Ikeda Sum Rule which
we discuss later.

Creation operators of QRPA phonons are defined by
\begin{equation}\label{co}
A^{\dagger}_{\omega}(\mu)=\sum_{pn}[X^{pn}_{\omega}(\mu)a^{\dagger}_{p}a^{\dagger}_{n}-Y^{pn}_{\omega}(\mu)a_{n}a_{\overline{p}}].
\end{equation}
In Eq.~\ref{co},  $(a^{\dagger}_{p(n)}, a_{p(n)})$ forms a
quasi-particle (q.p.) basis using Bogoliubov transformation
($\overline{p}$ represents time-reversed state of $p$), indices $p$
and $n$ denote $m_{p}\alpha_{p}$ and $m_{n}\alpha_{n}$,
respectively, and differentiate between proton and neutron
single-(quasi)-particle states. The sum runs over proton-neutron
pairs which satisfy $\mu=m_{p}-m_{n}$ and $\pi_{p}.\pi_{n}$=1, with
$\pi$ being parity, $m$ specifies the Nilsson eigenstates and
$\alpha$ represents additional quantum numbers.

For the separable forces, the pn-QRPA matrix equation can be
articulated more clearly as
\begin{equation}\label{x}
\begin{split}
X^{pn}_{\omega}=\frac{1}{\omega-\varepsilon_{pn}}[2\chi(q_{pn}Z^{-}_{\omega}+\tilde{q_{pn}}Z^{+}_\omega)\\
-2\kappa(q^{U}_{pn}Z^{- -}_{\omega}+q^{V}_{pn}Z^{+ +}_{\omega})],
\end{split}
\end{equation}
\begin{equation}\label{y}
\begin{split}
Y^{pn}_{\omega}=\frac{1}{\omega+\varepsilon_{pn}}[2\chi(q_{pn}Z^{+}_{\omega}+\tilde{q_{pn}}Z^{-}_\omega)\\
+2\kappa(q^{U}_{pn}Z^{+ +}_{\omega}+q^{V}_{pn}Z^{- -}_{\omega})],
\end{split}
\end{equation}
where $\varepsilon_{pn}=\varepsilon_{p}+\varepsilon_{n}$
(quasiparticle energies determined from a BCS calculation),\\
$q_{pn}=f_{pn}u_pv_n$, $q_{pn}^{U}=f_{pn}u_pu_n$,\\
$\tilde q_{pn}=f_{pn}v_pu_n$, $q_{pn}^{V}=f_{pn}v_pv_n$,\\
with
\begin{equation}\label{f}
f_{pn}(\mu)=\sum_{j_{p}j_{n}}D^{m_{p}\alpha_{p}}_{j_{p}}D^{m_{n}\alpha_{n}}_{j_{n}}<j_{p}m_{p}\mid
t-\sigma_{\mu}\mid j_{n}m_{n}>,
\end{equation}
as the single-particle GT transition amplitudes defined in the
Nilsson basis.  $D$ is  the transformation matrix constructed from
the Nilsson eigenfunctions.

\begin{equation}\label{Z-}
Z^{-}_{\omega}= \sum_{pn}(X^{pn}_{\omega}q_{pn}-Y^{pn}_{\omega}\tilde q_{pn})\\
\end{equation}
\begin{equation}\label{Z+}
Z^{+}_{\omega}= \sum_{pn}(X^{pn}_{\omega}\tilde q_{pn}-Y^{pn}_{\omega}q_{pn})\\
\end{equation}
\begin{equation}\label{Z--}
Z^{- -}_{\omega}= \sum_{pn}(X^{pn}_{\omega}q^{U}_{pn}+Y^{pn}_{\omega}q^{V}_{pn})\\
\end{equation}
\begin{equation}\label{Z++}
Z^{+ +}_{\omega}=
\sum_{pn}(X^{pn}_{\omega}q^{V}_{pn}+Y^{pn}_{\omega}q^{U}_{pn}).
\end{equation}

Using the normalization condition of the phonon amplitudes
\begin{equation}\label{DM1}
\sum_{pn}[(X^{pn}_{\omega})^{2}-(Y^{pn}_{\omega})^{2}]=1,
\end{equation}
the absolute values were determined by inserting $Z_{\omega}'s$ into
Eq.~\ref{x} and Eq.~\ref{y}.

GT transition amplitudes from the QRPA ground state $|-\rangle$
(QRPA vacuum; $A_{\omega}(\mu)|-\rangle=0)$ to one-phonon states
$|\omega,\mu\rangle=A^{\dagger}_{\omega}(\mu)|-\rangle$ were
calculated as
\begin{equation}\label{DM2}
\langle \omega,\mu|\sigma_{\mu}\tau_{\pm}|-\rangle=\mp
Z^{\pm}_{\omega}.
\end{equation}
Excitation energies of the one-phonon states are given by
$\omega-(\varepsilon_{p}+\varepsilon_{n})$, where $\varepsilon_{p}$
and $\varepsilon_{n}$ are energies of the single q.p. states of the
smallest q.p. energy in the proton and neutron systems,
respectively.

The RPA equation was solved for excitations from the $J^{\pi}=0^{+}$
ground state of an even-even nucleus. In the present model, excited
states of even-even nuclei were obtained by one-proton (or
one-neutron) excitations. They were described, in the q.p. picture,
by adding two-proton (two-neutron) q.p.'s to the ground state
\citep{mut92}. Transitions from these initial states were then
possible to final proton-neutron q.p. pair states in the odd-odd
daughter nucleus. The transition amplitudes and their reduction to
correlated ($c$) one-q.p. states are given by
\begin{eqnarray}
<p^{f}n_{c}^f \mid \sigma_{-\mu}\tau_{\pm} \mid p_{1}^{i}p_{2c}^{i}> \nonumber \\
 = -\delta (p^{f},p_{2}^{i}) <n_{c}^{f} \mid \sigma_{-\mu}\tau_{\pm} \mid
 p_{1c}^{i}> \nonumber \\
+\delta (p^{f},p_{1}^{i}) <n_{c}^{f} \mid \sigma_{-\mu}\tau_{\pm}
\mid p_{2c}^{i}>
 \label{first}
\end{eqnarray}
\begin{eqnarray}
\begin{split}
<p^{f}n_{c}^f \mid \sigma_{\mu}\tau_{\pm} \mid n_{1}^{i}n_{2c}^{i}> \nonumber \\
 = +\delta (n^{f},n_{2}^{i}) <p_{c}^{f} \mid \sigma_{\mu}\tau_{\pm} \mid
 n_{1c}^{i}> \nonumber \\
-\delta (n^{f},n_{1}^{i}) <p_{c}^{f} \mid \sigma_{\mu}\tau_{\pm}
\mid n_{2c}^{i}>
\end{split}
\end{eqnarray}
where $\mu$ = -1, 0, 1, are the spherical components of the spin
operator. For the construction of odd-A and odd-odd nuclei we refer
to \citep{mut92}.

Ikeda Sum Rule \citep{Ike64} is a mathematical tool which connects
the nucleon numbers (neutrons and protons) with the microscopic
structure of a given nucleus through the total Gamow Teller
strengths in either direction. It is formulated as:
\begin{equation}
{S^{+} -S^{-}} ={\sum _{f}\left\langle f\left|\sigma \tau_{-}
\right|i\right\rangle ^{2} - \sum _{f}\left\langle f\left|\sigma
\tau_{+} \right|i\right\rangle ^{2}}  ={3(N-Z)}\\
\end{equation}
This rule is model independent and provides a guideline for all
theoretical calculations of GT strength function. Theoretical
calculations of GT strength functions are supposed to fulfill the
Ikeda sum rule. In our model  the Hamiltonian (Eq.~\ref{Eqt. 9}) was
tuned to reproduce the Ikeda sum rule. In other words the daughter
spectrum was forced not to have a deficit of GT strength value.

The other requisite parameters for calculation of weak rates and
associated neutrino cooling and gamma ray heating rates are the
Nilsson potential parameters, the pairing gaps, the nuclear
quadrupole deformation, and the Q-values of the reactions.
Nilsson-potential parameters were adopted from Ref. \citep{Nil55}
and the Nilsson oscillator constant was chosen as $\hbar
\omega=41A^{-1/3}(MeV)$ (the same for protons and neutrons). The
relation for pairing gaps used in the present work is given as:
$\Delta _{p} =\Delta _{n} =12/\sqrt{A} (MeV)$. Nuclear quadrupole
deformation parameter play a significant role in the pn-QRPA
calculation, as argued by \citep{Sta90} and later by \citep{Nabi17}.
Measured (or theoretically predicted) values for the deformation
parameter $\beta$ were taken from \citep{Ram01}. Here the authors
estimated the $\beta$ parameter, for even-even isotopes, by relating
reduced electric quadrupole transition probability, B(E2)$\uparrow$
with the quadrupole deformation as:
\begin{equation}
\beta = \frac{4\pi}{3ZR_{0}^{2}}[B(E2)\uparrow/e^{2}]^{1/2}
\end{equation}
where $Z$ and $A$ are the atomic and mass numbers, respectively, and
$R_{0}^{2}=0.0144A^{2/3}$ barn. For remaining cases (non even-even),
the electric quadrupole deformation values were taken from
M\"{o}ller and collaborators \citep{Mol16}. Q-values were taken from
the recent mass compilation of Wang and collaborators
\citep{Wang12}. Using these new compilations and a model space of
5$\hbar\omega$, we calculated neutrino cooling  and gamma ray
heating rates associated with weak interaction processes in
astrophysical environments.
\subsection{The Rates Formalism:}
The basic formalism for calculation of stellar weak rates was
adopted from the pioneering work of Fuller, Fowler and Newman
\citep{Ful80,Ful82,Ful82a,Ful85}. However it is to be noted that,
unlike the work of Fuller, Fowler and Newman, we calculated all GT
strength distributions from parent excited states in a microscopic
fashion.

The weak decay rate from the \textit{ith} state (parent nucleus) to
the \textit{jth} state (daughter nucleus) is given by
\begin{equation}
\lambda _{ij} = \left(\frac{ln 2}{D} \right) [\phi_{ij} (T, \rho,
E_{f})][B(F)_{ij} + (g_{A}/ g_{V})^{2} B(GT)_{ij}]. \label{erates}
\end{equation}
The value of D was taken to be 6143s \citep{Har09}. $B(F)_{ij}$
denote the total reduced transition probabilities due to Fermi
interaction and $B(GT)_{ij}$ due to Gamow-Teller interactions. The
value of $(g_{A}/g_{V})$, denoting the ratio of axial and vector
coupling constants, was taken as -1.2694 \citep{Nak10}. The
$\phi_{ij}$ are the phase space integrals and are functions of
stellar temperature ($T$), density ($\rho$) and Fermi energy
($E_{f}$) of the electrons. The phase space integrals were
separately calculated using the relation (we use natural units in
this section, $\hbar=m_{e}=c=1$):
\begin{equation}
\phi_{ij} \, =\, \int _{1 }^{w_{m}}w\sqrt{w^{2} -1} (w_{m} \,
 -\, w)^{2} F(\pm Z,w)(1- D_{\mp}) dw,
\label{ps_ecap}
\end{equation}
whereas the phase space for continuum positron \emph{(lower signs)}
or electron \emph{(upper signs)} capture was calculated using:
\begin{equation}
\phi_{ij} \, =\, \int _{w_{l} }^{\infty }w\sqrt{w^{2} -1} (w_{m} \,
 +\, w)^{2} F(\pm Z,w)D_{\mp} dw.
\label{pscecap}
\end{equation}
In Eqs. ~(\ref{ps_ecap}) and ~(\ref{pscecap})  $w$ is the total
energy (rest+kinetic) of the electron or positron, $w_{l}$ is the
total threshold energy (rest+kinetic) for positron (or electron)
capture. F($ \pm$ Z,w) are the Fermi functions and were calculated
according to the procedure adopted by  \citep{Gov71}. Finite nuclear
size corrections as well as screening corrections were undertaken
within the recipe adopted by \citep{Gov71}. D$_{\pm}$ is the
Fermi-Dirac distribution function for positrons (electrons).
\begin{equation}
D_{+} =\left[\exp \left(\frac{E+2+E_{f} }{kT}\right)+1\right]^{-1},
\label{Gp}
\end{equation}
\begin{equation}
 D_{-} =\left[\exp \left(\frac{E-E_{f} }{kT}
 \right)+1\right]^{-1},
\label{Gm}
\end{equation}
where $E$ is the kinetic energy of the electrons and $k$ is the
Boltzmann constant.

The number density of protons associated with electrons and nuclei
is $\rho Y_{e} N_{A}$, where $\rho$ is the baryon density, $Y_{e}$
is the ratio of electron number to the baryon number, and $N_{A}$ is
the Avogadro number.
\begin{equation}\label{ye}
\rho Y_{e} = \frac{1}{\pi^{2}N_{A}}(\frac {m_{e}c}{\hbar})^{3}
\int_{0}^{\infty} (D_{-}-D_{+}) p^{2}dp,
\end{equation}
where $p=(w^{2}-1)^{1/2}$ is electron's momentum, and Eq.~\ref{ye}
has units of \textit{moles $cm^{-3}$}. This equation was used for an
iterative calculation of Fermi energies for selected values of $\rho
Y_{e}$ and $T$.

The calculation of neutrino and antineutrino cooling rates was
carried out by using the following equations:
\begin{equation}
\lambda ^{^{\nu(\bar{\nu})}} _{ij} = \left(\frac{ln 2}{D} \right)
[\phi_{ij}^{\nu} (T, \rho, E_{f})][B(F)_{ij} + (g_{A}/ g_{V})^{2}
B(GT)_{ij}]. \label{nurates}
\end{equation}
where all the constants and variables have their usual meaning as
discussed earlier in Eq. ~(\ref{erates}). The $\phi_{ij}^{\nu}$ are
the phase space integrals and are functions of stellar temperature
($T$), density ($\rho$) and Fermi energy ($E_{f}$) of the electrons.
They are explicitly given by
\begin{equation}
\phi_{ij}^{\nu} \, =\, \int _{1 }^{w_{m}}w\sqrt{w^{2} -1} (w_{m} \,
 -\, w)^{3} F(\pm Z,w)(1- D_{\mp}) dw,
\label{psedecay}
\end{equation}
and by
\begin{equation}
\phi_{ij}^{\nu} \, =\, \int _{w_{l} }^{\infty }w\sqrt{w^{2} -1}
(w_{m} \,
 +\, w)^{3} F(\pm Z,w)D_{\mp} dw.
\label{psecap}
\end{equation}
All symbols have meanings as discussed earlier.  For details on
calculation of reduced transition probabilities we refer to
\citep{Nabi04}. The total neutrino cooling rate per unit time per
nucleus is given by
\begin{equation}
\lambda^{\nu} =\sum _{ij}P_{i} \lambda _{ij}^{\nu}, \label{nurate}
\end{equation}
where $\lambda_{ij}^{\nu}$ is the sum of the electron capture and
positron decay rates for the transition $i \rightarrow j$.

The total gamma ray heating rates for a given nuclear specie are
given by:
\begin{equation}
\lambda^{\gamma} =\sum _{ij}P_{i} \lambda _{ij}E_{j}, \label{grate}
\end{equation}
where $P_{i}$, in Eqs. ~(\ref{nurate}) and  ~(\ref{grate}), denotes
the probability of occupation of parent excited states and follows
the Boltzmann statistical distribution, $\lambda_{ij}$ gives the sum
of the electron capture and positron decay rates,
$\lambda_{ij}=\lambda_{ij}^{ec}+\lambda_{ij}^{pd}$, or the sum of
positron capture and electron decay rates,
$\lambda_{ij}=\lambda_{ij}^{pc}+\lambda_{ij}^{ed}$, for the
transition $i \rightarrow j$ and E$_{j}$ is the energy of daughter's
excited state.

\section{Results and Discussion}

\begin{figure*}[h]
\includegraphics[scale=0.52]{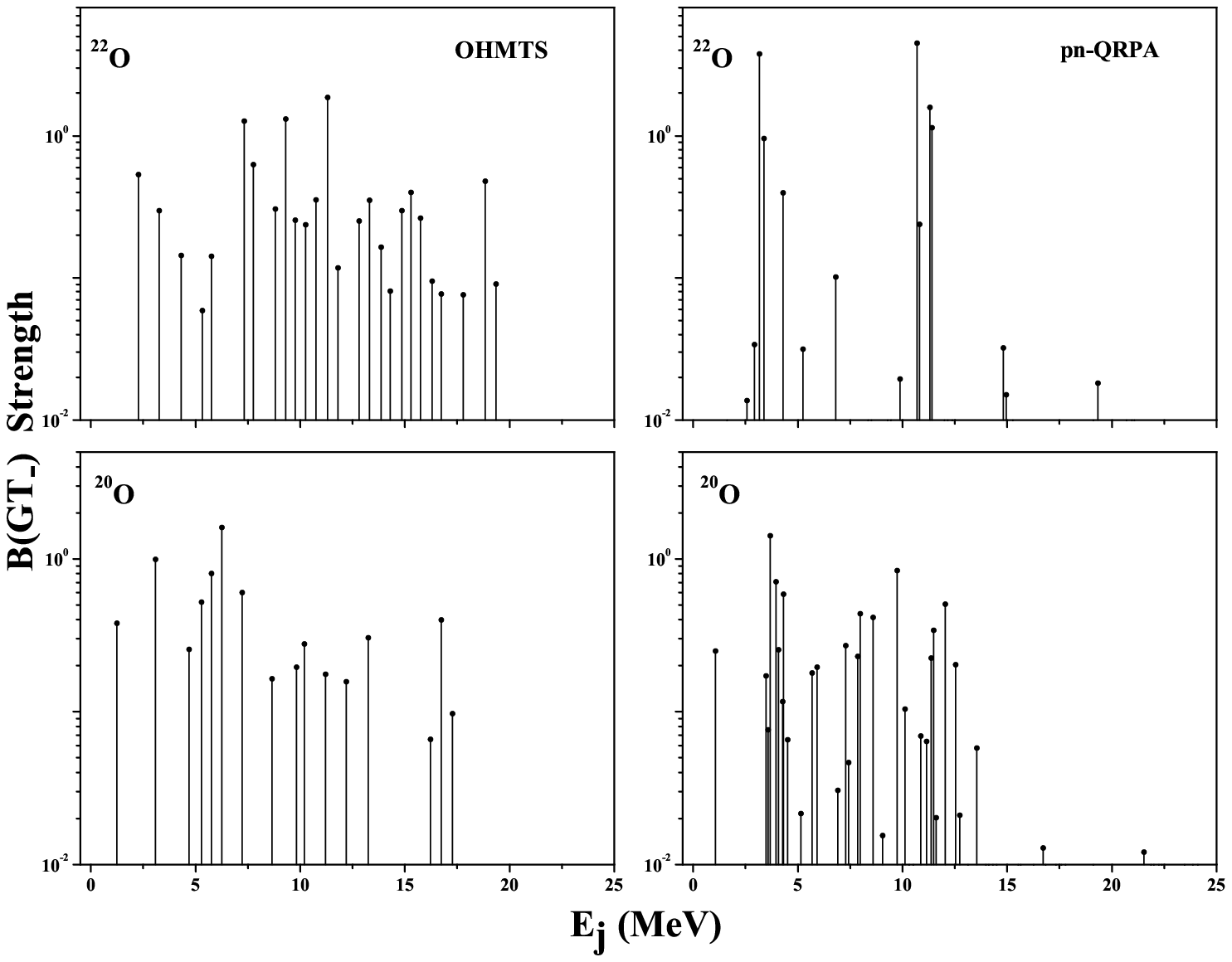}
\caption{Calculated B(GT) strength along $\beta$-decay direction for
even-even oxygen isotopes. The left panels show the \citep{Oda94}
calculation. Right panels show our calculated strength.}
\label{figa}
\end{figure*}

\begin{figure*}[h]
\includegraphics[scale=0.52]{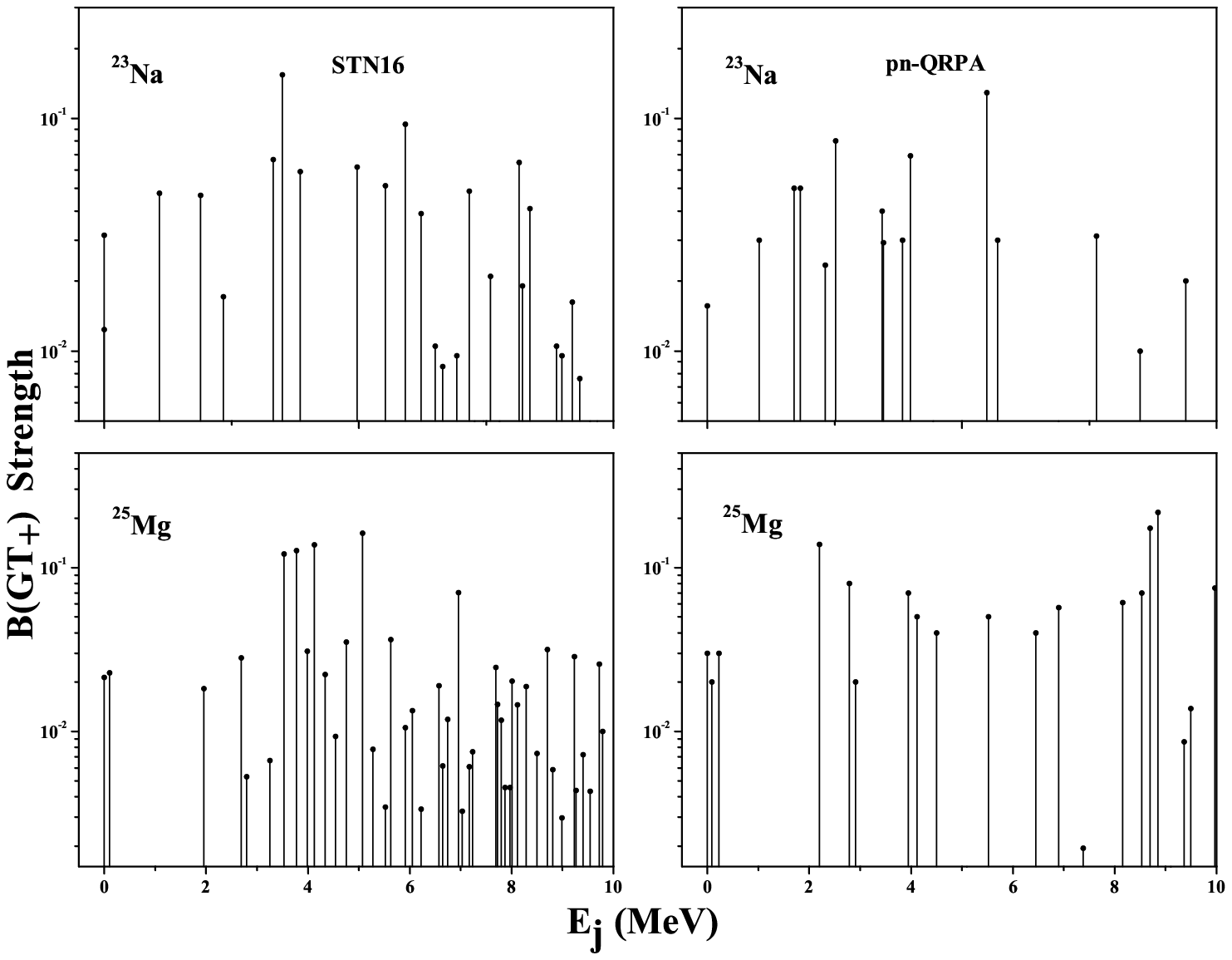}
\caption{Calculated B(GT) strength along electron capture direction
for odd-A nuclei. The left panels show the \citep{Suz16}
calculation. Right panels show our calculated strength.}
\label{figb}
\end{figure*}

Our calculated B(GT) strength for even-even and odd-A nuclei are
shown in Figs.~(\ref{figa}~-~\ref{figb}). Fig.~\ref{figa} shows the
pn-QRPA calculated B(GT) strength for even-even oxygen isotopes
along the $\beta$-decay direction. We also compare our calculated GT
strength with the shell model results of \citep{Oda94}.  For odd-A
nuclei we calculated our B(GT) strength values along electron
capture direction for $^{23}$Na and $^{25}$Mg and compared with the
shell model calculation of \citep{Suz16} (see Fig.~\ref{figb}).
Authors in \citep{Suz16} calculated GT strength distribution only
till 10 MeV in daughter nuclei. We also present our calculated B(GT)
strength distribution only up to excitation energy of 10 MeV  in
Fig.~\ref{figb}. The comparison of total GT strength and centroid
values of data shown in Figs~(\ref{figa}~-~\ref{figb}) is given in
Table~\ref{Table 0}. It may be seen from Table~\ref{Table 0} that
our model calculates bigger total GT strength and low residing
centroid values for even-even nuclei when compared with the shell
model results.  The comparison is interesting for odd-A nuclei. The
calculated GT strength distributions later directly affect the
computed weak rates in stellar matter.

\begin{figure*}[h]
\includegraphics[scale=0.52]{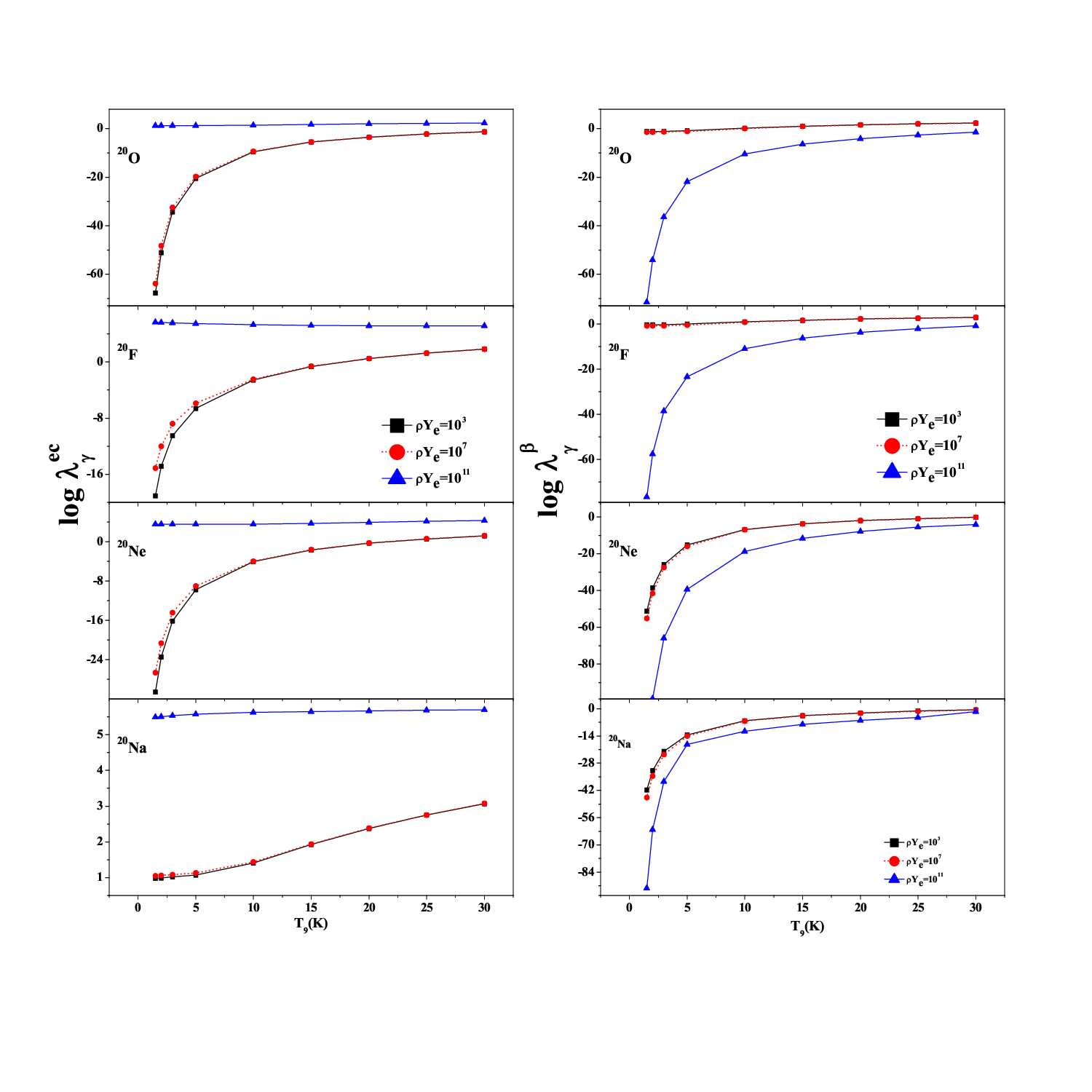}
\caption{Gamma ray heating rates due to $\beta$-decay and electron
capture processes on A = 20 nuclei, at selected stellar densities
$\rho$Y$_{e}$ (g.cm$^{-3}$) and temperatures. T$_{9}$ gives the
stellar temperature in units of $10^9$\;K.} \label{fig1}
\end{figure*}

\begin{figure*}[h]
\includegraphics[scale=0.52]{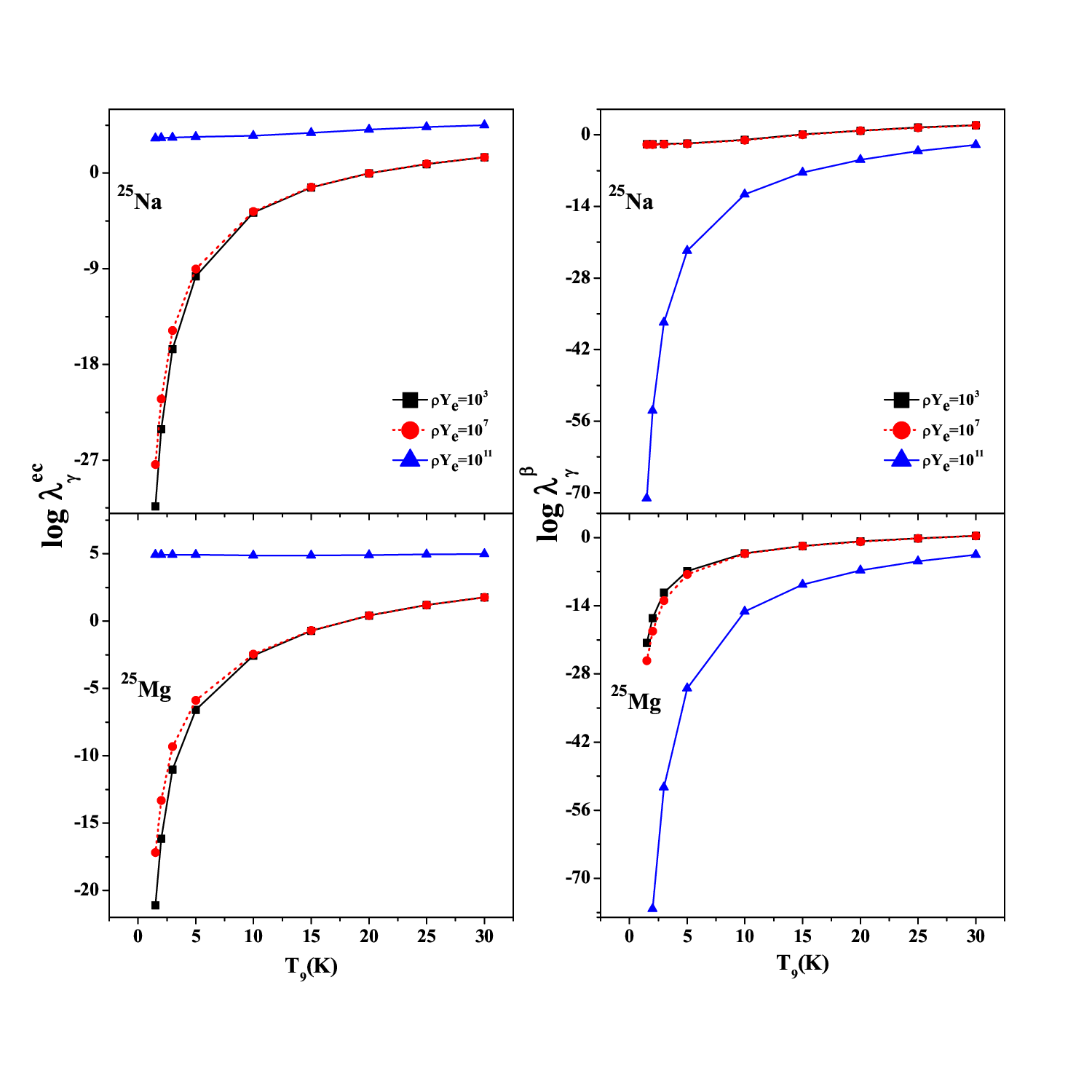}
\caption{Same as Fig.~\ref{fig1} but for A = 25 nuclei.}
\label{fig2}
\end{figure*}

The $\gamma$-ray heating rates due to A=20 nuclei are shown in
Fig.~\ref{fig1}, the left hand panel shows these rates in electron
capture direction and right hand panel in $\beta$-decay direction.
We see that with exception of $^{20}$O, where both rates are nearly
equal, the other nuclei like $^{20}$F, $^{20}$Ne and $^{20}$Na have
enhanced gamma heating rates due to electron capture reactions.
Specially, at high densities there is an orders of magnitude
increase in the gamma ray heating rates due to electron capture. A
similar trend may be seen for $^{25}$Mg and $^{25}$Na in
Fig.~\ref{fig2}.

\begin{figure*}[h]
\includegraphics[scale=0.52]{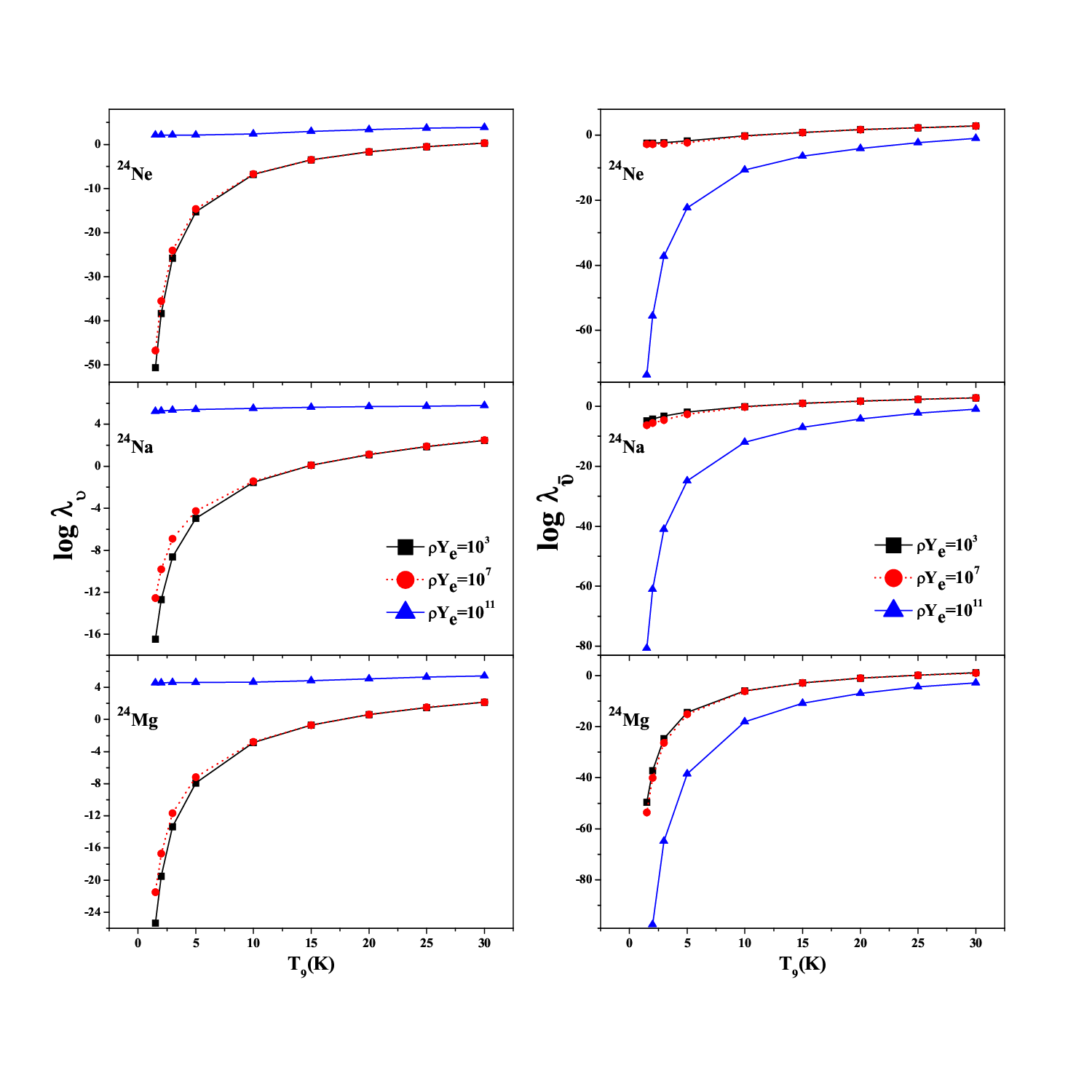}
\caption{Neutrino and antineutrino cooling rates due to
$\beta$-decay and electron capture processes on A = 24 nuclei, at
selected stellar densities $\rho$Y$_{e}$ (g.cm$^{-3}$) and
temperatures. T$_{9}$ gives the stellar temperature in units of
$10^9$\;K.} \label{fig3}
\end{figure*}

\begin{figure*}[h]
\includegraphics[scale=0.52]{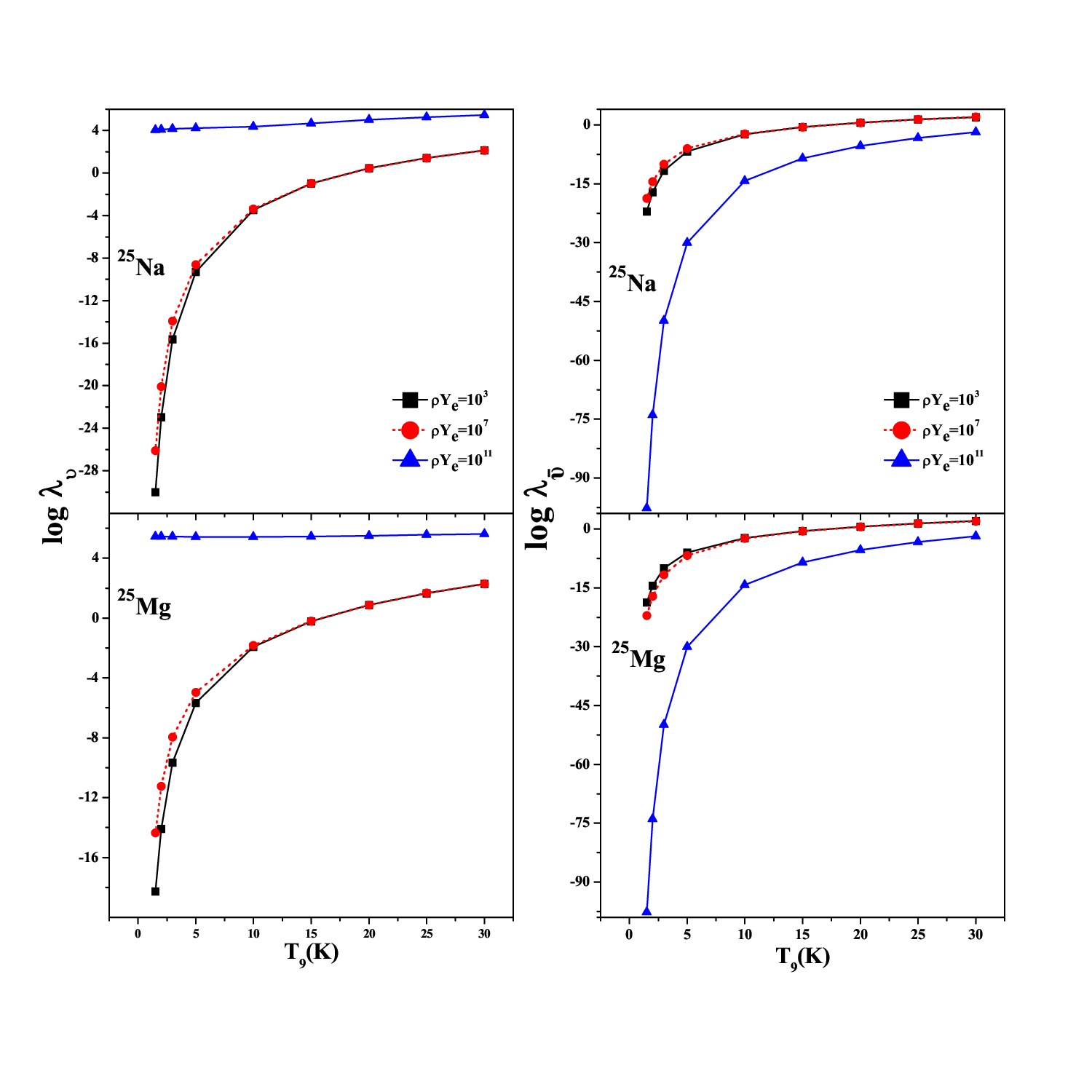}
\caption{Same as Fig.~\ref{fig3} but for A = 25 nuclei.}
\label{fig4}
\end{figure*}

The neutrino and antineutrino cooling rates due to weak rates on
A=24 nuclei can be seen in Fig.~\ref{fig3}. Here we have plotted the
rates for only $^{24}$Ne, $^{24}$Na and $^{24}$Mg at selected
densities. Logically, the neutrino cooling rates are orders of
magnitude more efficient than the corresponding  antineutrino rates,
though both escape the core without attenuation and tend to cool the
core. Fig.~\ref{fig4} shows the neutrino and antineutrino cooling
rates due to A=25 nuclei. For $^{25}$Na and $^{25}$Mg, a similar
trend is witnessed for the cooling rates.

We display our calculated heating and cooling rates for  selected
$sd$ shell nuclei in Tables~(\ref{Table 1}~-~\ref{Table 7}).
Table~\ref{Table 1} shows the associated gamma ray heating rates and
Table~\ref{Table 2} the associated neutrino (antineutrino) cooling
rates for $^{20}$O, $^{20}$F, $^{20}$Na and $^{20}$Na, respectively,
as a function of stellar temperature and density. Table~\ref{Table
3} and Table~\ref{Table 4} show  results for $^{23}$O, $^{23}$F,
$^{23}$Ne, $^{23}$Na and $^{23}$Mg in a similar format.
Table~\ref{Table 5} and Table~\ref{Table 6} enlist the gamma ray
heating rates and neutrino cooling rates for A=24 nuclear species,
respectively. Table~\ref{Table 7} shows the pn-QRPA calculated
heating and cooling rates for $^{25}$Na and $^{25}$Mg. From these
tables, it can be concluded, that the gamma ray heating and neutrino
cooling rates, due to both electron capture and $\beta$-decay on
$sd$ shell nuclei, increase with increasing core temperature. This
is because of a corresponding increment in the electron capture and
$\beta$-decay rates on these nuclei with increasing stellar
temperature. The $\beta$-decay rates decrease with increasing
stellar density (because of increase in Fermi energy and decrease in
phase space factor). Accordingly we note that the heating rate along
$\beta$-decay direction decreases with increasing density. A similar
explanation follows for the neutrino cooling rates. We note that the
heating and cooling rates play a dominant role in the stellar core
evolution of 8-10 M$_\odot$ stars. The ASCII files of all calculated
heating and cooling rates are available and may be requested from
corresponding author to be used in simulation codes.

We compare our current calculation with previous published work of
\citep{Nabi99}(referred to as ADNDT99) using the same theory. We
also compare our work with previous shell model calculations:
\citep{Oda94} (referred to as OHMTS) and \citep{Suz16} (referred to
as STN16). It is to be noted that the shell model calculation of
STN16 includes the effects of Coulomb corrections while those of
OHMTS do not. The electron capture, neutrino cooling and gamma
heating rates are generally smaller for STN16 when compared with
OHMTS rates. We also incorporated Coulomb corrections in our
calculation.

The mutual comparison may be seen in Tables~(\ref{Table
8}~-~\ref{Table 9}) for A = 23 and Tables~(\ref{Table
10}~-~\ref{Table 11}) for A = 25. For the comparison we selected the
pairs $^{23}$Na $\rightarrow^{23}$Ne, $^{23}$Ne $\rightarrow^{23}$F
and $^{25}$Mg $\rightarrow^{25}$Na, $^{25}$Na $\rightarrow^{25}$Ne.
All these tables are shown in the electron capture direction. In
these tables the right hand column (group of four sub columns)
enlists the respective neutrino cooling rates and a similar column
on the left hand shows the gamma ray heating rates in each table.

Comparing the current gamma heating rates due to electron capture
rates on $^{23}$Na with ADNDT99 rates (Table~\ref{Table 8}), we note
that current rates are up to an order of magnitude bigger at high
temperatures. The new gamma heating rates are in better comparison
with shell model rates, specially at high temperatures and
densities. For  density range $10^{10}-10^{11}$ (g.cm$^{-3}$) the
comparison is fair albeit the shell model rates are factor 3--8
bigger. At T$_{9}$ = 30, the shell model rates are factor 3--4
bigger. The new calculated neutrino cooling rates, on the other
hand, are generally in better agreement with shell model rates.

Table~\ref{Table 9} shows that the new pn-QRPA calculated gamma
heating rates, due to electron capture on $^{23}$Ne, are up to an
order of magnitude bigger compared with ADNDT99 rates. The
comparison with ADNDT99 rates improves as stellar core stiffens to
high densities. It is noted that at low temperatures, T$_{9}$ =
0.01--0.1, shell model gamma heating rates are many orders of
magnitude smaller than pn-QRPA rates. Only at density $10^{11}$
(g.cm$^{-3}$) are the two rates in good agreement. This is because
the pn-QRPA model incorporated particle emission processes in their
calculation not considered in shell model calculations. In the
pn-QRPA calculation all excited states, with energy less than
separation energy of neutron (proton) decay directly to ground state
via $\gamma$ transitions. At high temperatures T$_{9}$ = 30, our
rates are in very good agreement with shell model rates. The pn-QRPA
calculated neutrino cooling rates are up to an order of magnitude
bigger than shell model rates.

The pn-QRPA calculated gamma heating rates, due to electron capture
on $^{25}$Mg, are in decent agreement with shell model rates for
high densities $10^{10}-10^{11}$ (g.cm$^{-3}$), as can be seen from
Table~\ref{Table 10}. For smaller density range the shell model
heating rates are orders of magnitude bigger (specially at low
temperatures). Our neutrino cooling rates are an order of magnitude
bigger than shell model rates. However at T$_{9}$ = 30, the pn-QRPA
and shell model rates are in good agreement.

Table~\ref{Table 11} shows that our gamma heating rates, due to
electron capture on $^{25}$Na, are in good concordance with shell
model rates for high densities $10^{10}-10^{11}$ (g.cm$^{-3}$).
Contrary to Table~\ref{Table 10}, at low temperatures and densities
the pn-QRPA gamma heating rates are orders of magnitude bigger than
shell model heating rates. The neutrino cooling rates are again an
order of magnitude bigger than shell model rates.

Finally we present the comparison of gamma heating and neutrino
cooling rates for a sample even-even nucleus. Table~\ref{Table 12}
compares the gamma heating and neutrino cooling rates due to
$\beta^{-}$-decay from $^{20}$O. It is again noted that
\citep{Nabi99} did not calculate weak rates for nuclei with Z $<$ 10
and as such we compare our calculation with the shell model
calculations of OHMTS and STN16. A very decent comparison between
pn-QRPA and shell model calculations is noted for the gamma heating
rates due to $\beta$-decay from $^{20}$O. Only at T$_{9}$ = 30 are
the shell model rates factor four bigger. OHMTS also considered
transitions from excited states much higher than the lesser of the
$S_p$ or $S_n$ values, which causes their rates to increase at
higher temperatures. We argue that such contributions should not be
considered. At any $E_i$ higher than the lesser of $S_p$ or $S_n$
(after accounting for the effective Coulomb barrier hindering the
emission of prompt protons and for the uncertainty in the
calculation of energy levels), the nucleus will emit protons or
neutrons instead of continuing to undergo $\beta$-decay. On the
other hand the pn-QRPA calculated neutrino cooling rates are orders
of magnitude smaller, specially at low temperatures. At T$_{9}$ =
30, the pn-QRPA cooling rates are an order of magnitude smaller. It
should be pointed out that the  rates are sensitive functions of the
difference of parent and daughter energies ($E_i - E_j$) and just by
a mere addition of around 0.5 MeV, many of our small rates could
increase by orders of magnitude.

To summarize the comparison of current work with previous
calculations, it can be noted that the new pn-QRPA rates are up to
an order of magnitude bigger than ADNDT99 rates. This we attribute
to optimal choice of model parameters and use of latest values of
mass defect and physical constants. Shell model rates are generally
in excellent agreement with our calculation at high temperatures. At
low temperatures and densities, orders of magnitude difference is
witnessed. This is due to multiple effects. Firstly shell model did
not take into account particle emission processes from parent
excited states. The calculation of energy eigenvalues in the two
models may contribute to the differences in the total rates (see
Eqs. ~(\ref{nurate}) and ~(\ref{grate})). But also the calculation
of total GT strength values and placement of GT centroids, in the
two models, resulted in this difference. The pn-QRPA model generally
favors cooler cores due to bigger cooling rates (see also discussion
in \citep{Nabi99}). It is now left to the simulators to analyze
these differences and check whether their prediction makes the core
evolve towards ECSN, CCSN or any other interesting occurrence.


\section{Conclusions}
\label{sec:conclusions}

Semi-leptonic weak-interaction mediated rates for nuclei in stellar
core are important factors in determining the thermodynamic and
hydrodynamic trajectories of stars in the late stages of their
evolution. The outcome of rapid contraction of electron-degenerate
O-Ne-Mg cores of 8-10 M$_\odot$ stars is determined by contraction,
heating and cooling processes in the stellar core. All these
processes compete with one another. Result of these competitive
processes decide whether the star would end up its life as a
core-collapse or electron capture supernova. In this work we
presented gamma ray heating and (anti)neutrino cooling rates due to
electron capture and $\beta$-decays on $sd$-shell nuclei in stellar
environment. The cooling rates are important for the A = 23 and 25
nuclei for nuclear URCA process. On the other hand A = 20 and 24
nuclei rates are significant for heat generation and
core-contraction rates in the O-Ne-Mg cores.

For our calculation we used the pn-QRPA in a multi-shell deformed
single-particle space with a schematic separable interaction to
calculate gamma heating and (anti)neutrino cooling rates due to weak
rates on key $sd$-shell nuclei. An optimum choice of model
parameters and  incorporation of recent experimental mass
compilation resulted in a reliable calculation of Gamow-Teller
strength distributions that fulfilled the model-independent Ikeda
sum rule. The cooling and heating rates were calculated for density
range ($10 \leq \rho($\;g.cm$^{-3}) \leq $ 10$^{11}$) and
temperature range ($0.01\times10^{9}$ $\leq$ $\;T(K)$ $\leq$
$30\times10^{9}$). The calculated rates were  compared with previous
pn-QRPA calculation (ADNDT99) as well as previous shell model rates
(OHMTS and STN16) and interesting differences were reported.

The pn-QRPA calculated gamma heating rates are generally bigger than
those calculated by shell model rates. $^{25}$Mg was the only
exception where at density range $10^{7}-10^{9}$ (g.cm$^{-3}$) the
shell model heating rates were orders of magnitude bigger. At high
temperatures the shell model and pn-QRPA gamma heating rates are in
reasonable agreement. At times the shell model rates are slightly
bigger and sometimes the pn-QRPA heating rates are marginally
bigger. The neutrino cooling rates calculated using the pn-QRPA
model is up to an order of magnitude bigger compared with shell
model rates for the A = 23 and 25 nuclei (odd-A cases). The
implications of these cooling and heating rates worth attention of
core-collapse simulators. They are urged to test-run our calculated
rates to see whether the stars end up as a white dwarf, an ECSN or
CCSN.

\acknowledgments  The authors would like to acknowledge the
assistance of the anonymous referee which led to a substantial
improvement of the paper. J.-U. Nabi would like to acknowledge the
support of the Higher Education Commission Pakistan through the HEC
Projects No. 20-3099 and 5557/KPK/NRPU/R$\&$D/HEC/2016.

\clearpage \onecolumn

\clearpage
\begin{table*}[!htbp]
\centering\caption{Comparison of calculated total GT strength
(B(GT)$_{+}$) and centroid (E$_{+}$) values, along electron capture
direction, for $^{23}$Na and $^{25}$Mg and, along $\beta$-decay
direction, for $^{20}$O and $^{22}$O.}\label{Table 0}
\begin{tabular}{|c|cc|cc|}
\hline Nuclei & \multicolumn{2}{c|}{$\sum B(GT)_{+}$} &\multicolumn{2}{c|}{$E_{+}$ (MeV)} \\
\cline{2-5}
&STN16 & This Work &STN16 &This Work\\
\hline
$^{23}$Na & 0.964 & 0.64 & 4.99 &3.90\\
$^{25}$Mg& 1.19 & 1.25 & 5.25& 6.16\\
\hline & \multicolumn{2}{c|}{$\sum B(GT)_{-}$} &\multicolumn{2}{c|}{$E_{-}$ (MeV)} \\
\cline{2-5}&OHMTS & This Work &OHMTS &This Work\\
\hline $^{20}$O & 6.98 & 8.01 & 7.15 & 4.88\\
$^{22}$O & 10.14& 12.90 &10.36& 7.84\\
\hline
\end{tabular}
  \end{table*}
\begin{table*}[!htbp]
\centering \tiny \caption{Calculated $\gamma$-ray heating rates due
to beta decay ($\lambda_{\gamma}^{\beta}$) and electron capture
($\lambda_{\gamma}^{EC}$) on $^{20}$O, $^{20}$F, $^{20}$Ne and
$^{20}$Na at different selected densities and temperatures in
stellar environment. The first column shows the stellar densities
($\rho$Y$_{e}$) (in units of g.cm$^{-3}$). The rates are tabulated
in units of MeV.s$^{-1}$. In the table entry 1.00$\times10^{-100}$
means that calculated rate is smaller than 1.00$\times10^{-100}$
MeV.s$^{-1}$.}\label{Table 1}
\resizebox{\columnwidth}{!}{%
\begin{tabular}{|c|c|cc|cc|cc|cc|}
\hline $\rho$$\it Y_{e}$ & T (K)  &
\multicolumn{2}{c|}{$^{20}$O}&\multicolumn{2}{c|}{$^{20}$F}&
\multicolumn{2}{c|}{$^{20}$Ne} & \multicolumn{2}{c|}{$^{20}$Na}\\
\cline{3-10}(g.cm$^{-3}$) & &$\lambda_{\gamma}^{\beta}$ &
$\lambda_{\gamma}^{EC}$ & $\lambda_{\gamma}^{\beta}$ &
$\lambda_{\gamma}^{EC}$ &$\lambda_{\gamma}^{\beta}$ & $\lambda_{\gamma}^{EC}$&$\lambda_{\gamma}^{\beta}$ & $\lambda_{\gamma}^{EC}$\\
\hline
10$^{1}$ & 1.00$\times10^{7}$& 6.14$\times10^{-2}$& 1.00$\times10^{-100 }$& 4.13$\times10^{-1}$& 1.00$\times10^{-100 }$& 1.00$\times10^{-100 }$& 1.00$\times10^{-100 }$& 1.00$\times10^{-100 }$& 9.44$\times10^{0} $\\
 10$^{1}$& 4.00$\times10^{8}$& 6.14$\times10^{-2}$& 1.00$\times10^{-100 }$& 4.13$\times10^{-1}$& 6.22$\times10^{-60 }$& 1.00$\times10^{-100 }$& 1.00$\times10^{-100 }$& 1.00$\times10^{-100 }$& 9.44$\times10^{0}$\\
 10$^{1}$& 7.00$\times10^{8}$& 6.14$\times10^{-2}$& 1.00$\times10^{-100 }$& 4.13$\times10^{-1}$& 7.11$\times10^{-38 }$& 1.00$\times10^{-100 }$& 8.53$\times10^{-63 }$& 1.72$\times10^{-87 }$& 9.44$\times10^{0}$\\
 10$^{1}$& 1.00$\times10^{9}$& 6.14$\times10^{-2}$& 1.00$\times10^{-100 }$& 4.11$\times10^{-1}$& 7.31$\times10^{-28 }$& 3.85$\times10^{-77 }$& 3.05$\times10^{-45 }$& 3.29$\times10^{-62 }$& 9.44$\times10^{0}$\\
 10$^{1}$& 2.00$\times10^{9}$& 6.19$\times10^{-2}$& 7.82$\times10^{-52 }$& 3.93$\times10^{-1}$& 1.30$\times10^{-15 }$& 2.54$\times10^{-39 }$& 3.24$\times10^{-24 }$& 1.30$\times10^{-32 }$& 9.77$\times10^{0}$\\
 10$^{1}$& 3.00$\times10^{9}$& 6.65$\times10^{-2}$& 5.38$\times10^{-35 }$& 4.03$\times10^{-1}$& 3.06$\times10^{-11 }$& 1.57$\times10^{-26 }$& 7.48$\times10^{-17 }$& 1.29$\times10^{-22 }$& 1.04$\times10^{1}$\\
\hline
\end{tabular}%
}
\end{table*}

\begin{table*}[htb]
\centering \tiny\caption{Calculated antineutrino and neutrino
cooling rates due to weak rates on $^{20}$O, $^{20}$F, $^{20}$Ne and
$^{20}$Na at different selected densities and temperatures in
stellar environment. The first column shows the stellar densities
($\rho$Y$_{e}$) (in units of g.cm$^{-3}$). The rates are tabulated
in units of MeV.s$^{-1}$.}\label{Table 2} 
\resizebox{\columnwidth}{!}{%
\begin{tabular}{|c|c|cc|cc|cc|cc|}
\hline $\rho$$\it Y_{e}$ & T (K)  &
\multicolumn{2}{c|}{$^{20}$O}&\multicolumn{2}{c|}{$^{20}$F}&
\multicolumn{2}{c|}{$^{20}$Ne} & \multicolumn{2}{c|}{$^{20}$Na}\\
\cline{3-10}(g.cm$^{-3}$) & &$\lambda_{\bar\nu}$ & $\lambda_{\nu}$ &
$\lambda_{\bar\nu}$ & $\lambda_{\nu}$ &$\lambda_{\bar\nu}$ & $\lambda_{\nu}$&$\lambda_{\bar\nu}$ & $\lambda_{\nu}$\\
\hline
10$^{1}$& 1.00$\times10^{7}$& 7.94$\times10^{-2}$& 1.00$\times10^{-100 }$& 5.41$\times10^{-2}$& 1.00$\times10^{-100 }$& 1.00$\times10^{-100 }$& 1.00$\times10^{-100 }$& 1.00$\times10^{-100 }$& 4.81$\times10^{0}$\\
 10$^{1}$& 4.00$\times10^{8}$& 7.94$\times10^{-2}$& 1.00$\times10^{-100 }$& 5.41$\times10^{-2}$& 7.74$\times10^{-60 }$& 1.00$\times10^{-100 }$& 1.00$\times10^{-100 }$& 1.00$\times10^{-100 }$& 4.81$\times10^{0}$\\
 10$^{1}$& 7.00$\times10^{8}$& 7.94$\times10^{-2}$& 1.00$\times10^{-100 }$& 5.40$\times10^{-2}$& 1.51$\times10^{-37 }$& 1.00$\times10^{-100 }$& 5.05$\times10^{-63 }$& 1.52$\times10^{-83 }$& 4.81$\times10^{0}$\\
 10$^{1}$& 1.00$\times10^{9}$& 7.94$\times10^{-2}$& 1.37$\times10^{-98 }$& 5.38$\times10^{-2}$& 2.04$\times10^{-27 }$& 5.36$\times10^{-75 }$& 2.72$\times10^{-45 }$& 4.60$\times10^{-59 }$& 4.82$\times10^{0}$\\
 10$^{1}$& 2.00$\times10^{9}$& 8.04$\times10^{-2}$& 9.06$\times10^{-51 }$& 5.50$\times10^{-2}$& 5.56$\times10^{-15 }$& 3.58$\times10^{-38 }$& 5.78$\times10^{-24 }$& 2.32$\times10^{-30 }$& 4.99$\times10^{0}$\\
 10$^{1}$& 3.00$\times10^{9}$& 8.75$\times10^{-2}$& 2.00$\times10^{-34 }$& 8.05$\times10^{-2}$& 1.68$\times10^{-10 }$& 1.21$\times10^{-25 }$& 1.86$\times10^{-16 }$& 1.21$\times10^{-20 }$& 5.33$\times10^{0}$\\
\hline
\end{tabular}%
}
\end{table*}
\begin{table*}[htb]
\centering \tiny \caption{Same as Table.~\ref{Table 1}, but for
$^{23}$O, $^{23}$F, $^{23}$Ne, $^{23}$Na and $^{23}$Mg}\label{Table
3}
\resizebox{\columnwidth}{!}{%
\begin{tabular}{|c|c|cc|cc|cc|cc|cc|}
\hline $\rho$$\it Y_{e}$ & T (K)  &
\multicolumn{2}{c|}{$^{23}$O}&\multicolumn{2}{c|}{$^{23}$F}&
\multicolumn{2}{c|}{$^{23}$Ne} & \multicolumn{2}{c|}{$^{23}$Na} & \multicolumn{2}{c|}{$^{23}$Mg}\\
\cline{3-12}(g.cm$^{-3}$) & &$\lambda_{\gamma}^{\beta}$ &
$\lambda_{\gamma}^{EC}$ & $\lambda_{\gamma}^{\beta}$ &
$\lambda_{\gamma}^{EC}$ &$\lambda_{\gamma}^{\beta}$ & $\lambda_{\gamma}^{EC}$&$\lambda_{\gamma}^{\beta}$ & $\lambda_{\gamma}^{EC}$ &$\lambda_{\gamma}^{\beta}$ & $\lambda_{\gamma}^{EC}$\\
\hline
10$^{1}$& 1.00$\times10^{7}$& 4.44$\times10^{1}$& 1.00$\times10^{-100}$& 1.58$\times10^{0}$& 1.00$\times10^{-100 }$& 1.16$\times10^{-2}$& 1.00$\times10^{-100 }$& 1.00$\times10^{-100 }$& 1.00$\times10^{-100 }$& 1.00$\times10^{-100 }$& 4.79$\times10^{-2}$\\
 10$^{1}$& 4.00$\times10^{8}$& 4.44$\times10^{1}$& 1.00$\times10^{-100}$& 1.58$\times10^{0}$& 1.00$\times10^{-100 }$& 1.16$\times10^{-2}$& 1.00$\times10^{-100 }$& 4.28$\times10^{-70 }$& 6.68$\times10^{-67 }$& 1.00$\times10^{-100 }$& 4.19$\times10^{-2}$\\
 10$^{1}$& 7.00$\times10^{8}$& 4.44$\times10^{1}$& 1.00$\times10^{-100}$& 1.58$\times10^{0}$& 3.07$\times10^{-91 }$& 1.16$\times10^{-2}$& 8.02$\times10^{-71 }$& 9.12$\times10^{-45 }$& 3.52$\times10^{-42 }$& 1.00$\times10^{-100 }$& 3.61$\times10^{-2}$\\
 10$^{1}$& 1.00$\times10^{9}$& 4.44$\times10^{1}$& 1.00$\times10^{-100}$& 1.58$\times10^{0}$& 1.01$\times10^{-64 }$& 1.16$\times10^{-2}$& 8.34$\times10^{-51 }$& 2.62$\times10^{-32 }$& 3.73$\times10^{-31 }$& 1.81$\times10^{-75 }$& 3.29$\times10^{-2}$\\
 10$^{1}$& 2.00$\times10^{9}$& 4.30$\times10^{1}$& 4.60$\times10^{-71}$& 1.58$\times10^{0}$& 2.47$\times10^{-33 }$& 1.25$\times10^{-2}$& 8.65$\times10^{-27 }$& 1.26$\times10^{-17 }$& 9.33$\times10^{-18 }$& 4.63$\times10^{-43 }$& 3.02$\times10^{-2}$\\
\hline
\end{tabular}%
}
\end{table*}

\begin{table*}[htb]
\centering \tiny \caption{Same as Table.~\ref{Table 2}, but for
$^{23}$O, $^{23}$F, $^{23}$Ne, $^{23}$Na and $^{23}$Mg}\label{Table
4}
\resizebox{\columnwidth}{!}{%
\begin{tabular}{|c|c|cc|cc|cc|cc|cc|}
\hline $\rho$$\it Y_{e}$ & T (K)  &
\multicolumn{2}{c|}{$^{23}$O}&\multicolumn{2}{c|}{$^{23}$F}&
\multicolumn{2}{c|}{$^{23}$Ne} & \multicolumn{2}{c|}{$^{23}$Na} & \multicolumn{2}{c|}{$^{23}$Mg}\\
\cline{3-12}(g.cm$^{-3}$) & &$\lambda_{\bar\nu}$ & $\lambda_{\nu}$ &
$\lambda_{\bar\nu}$ & $\lambda_{\nu}$ &$\lambda_{\bar\nu}$ & $\lambda_{\nu}$&$\lambda_{\bar\nu}$ & $\lambda_{\nu}$ & $\lambda_{\bar\nu}$ & $\lambda_{\nu}$\\
\hline
10$^{1}$& 1.00$\times10^{7}$& 1.73$\times10^{1}$& 1.00$\times10^{-100 }$& 4.28$\times10^{-1}$& 1.00$\times10^{-100 }$& 4.05$\times10^{-2}$& 1.00$\times10^{-100 }$& 1.00$\times10^{-100 }$& 1.00$\times10^{-100 }$& 1.00$\times10^{-100 }$& 8.13$\times10^{-2}$\\
 10$^{1}$& 4.00$\times10^{8}$& 1.73$\times10^{1}$& 1.00$\times10^{-100 }$& 4.28$\times10^{-1}$& 1.00$\times10^{-100 }$& 4.05$\times10^{-2}$& 1.00$\times10^{-100 }$& 1.99$\times10^{-60 }$& 6.58$\times10^{-66 }$& 1.00$\times10^{-100 }$& 7.26$\times10^{-2}$\\
 10$^{1}$& 7.00$\times10^{8}$& 1.73$\times10^{1}$& 1.00$\times10^{-100 }$& 4.28$\times10^{-1}$& 3.21$\times10^{-89 }$& 4.05$\times10^{-2}$& 1.11$\times10^{-69 }$& 6.50$\times10^{-36 }$& 5.46$\times10^{-41 }$& 3.98$\times10^{-93 }$& 6.43$\times10^{-2}$\\
 10$^{1}$& 1.00$\times10^{9}$& 1.73$\times10^{1}$& 1.00$\times10^{-100 }$& 4.28$\times10^{-1}$& 4.19$\times10^{-63 }$& 4.06$\times10^{-2}$& 1.13$\times10^{-49 }$& 2.44$\times10^{-26 }$& 8.02$\times10^{-30 }$& 2.03$\times10^{-65 }$& 5.97$\times10^{-2}$\\
 10$^{1}$& 2.00$\times10^{9}$& 1.68$\times10^{1}$& 1.57$\times10^{-77 }$& 4.31$\times10^{-1}$& 4.91$\times10^{-32 }$& 4.27$\times10^{-2}$& 9.91$\times10^{-26 }$& 1.67$\times10^{-14 }$& 3.09$\times10^{-16 }$& 5.83$\times10^{-33 }$& 5.69$\times10^{-2}$\\
 10$^{1}$& 3.00$\times10^{9}$& 1.60$\times10^{1}$& 6.07$\times10^{-52 }$& 4.54$\times10^{-1}$& 2.49$\times10^{-21 }$& 5.36$\times10^{-2}$& 2.09$\times10^{-17 }$& 2.20$\times10^{-10 }$& 2.19$\times10^{-11 }$& 5.61$\times10^{-22 }$& 7.41$\times10^{-2}$\\
\hline
\end{tabular}%
}
\end{table*}
\begin{table*}[htb]
\centering \scriptsize \caption{Same as Table.~\ref{Table 1}, but
for $^{24}$Ne, $^{24}$Na and $^{24}$Mg.}\label{Table 5}
\resizebox{\columnwidth}{!}{%
\begin{tabular}{|c|c|cc|cc|cc|}
\hline $\rho$$\it Y_{e}$ & T (K) & \multicolumn{2}{c|}{$^{24}$Ne}&
\multicolumn{2}{c|}{$^{24}$Na} & \multicolumn{2}{c|}{$^{24}$Mg}\\
\cline{3-8}(g.cm$^{-3}$) & &$\lambda_{\gamma}^{\beta}$ &
$\lambda_{\gamma}^{EC}$ & $\lambda_{\gamma}^{\beta}$ &
$\lambda_{\gamma}^{EC}$ &$\lambda_{\gamma}^{\beta}$ & $\lambda_{\gamma}^{EC}$\\
\hline
10$^{1}$& 1.00$\times10^{7}$& 2.31$\times10^{-3}$& 1.00$\times10^{-100 }$& 6.78$\times10^{-5}$& 1.00$\times10^{-100 }$& 1.00$\times10^{-100 }$& 1.00$\times10^{-100 }$\\
 10$^{1}$& 4.00$\times10^{8}$& 2.31$\times10^{-3}$& 1.00$\times10^{-100 }$& 6.78$\times10^{-5}$& 9.86$\times10^{-52 }$& 1.00$\times10^{-100 }$& 1.32$\times10^{-83 }$\\
 10$^{1}$& 7.00$\times10^{8}$& 2.31$\times10^{-3}$& 1.00$\times10^{-100 }$& 6.89$\times10^{-5}$& 1.08$\times10^{-32 }$& 1.00$\times10^{-100 }$& 1.47$\times10^{-51 }$\\
 10$^{1}$& 1.00$\times10^{9}$& 2.31$\times10^{-3}$& 8.36$\times10^{-77 }$& 7.52$\times10^{-5}$& 6.28$\times10^{-24 }$& 3.19$\times10^{-76 }$& 1.44$\times10^{-37 }$\\
 10$^{1}$& 2.00$\times10^{9}$& 2.36$\times10^{-3}$& 5.15$\times10^{-40 }$& 2.84$\times10^{-4}$& 3.59$\times10^{-13 }$& 5.78$\times10^{-39 }$& 1.08$\times10^{-20 }$\\
 10$^{1}$& 3.00$\times10^{9}$& 2.81$\times10^{-3}$& 2.07$\times10^{-27 }$& 1.82$\times10^{-3}$& 2.88$\times10^{-9}$& 2.16$\times10^{-26 }$& 1.21$\times10^{-14 }$\\
\hline
\end{tabular}%
}
\end{table*}
\begin{table*}[htb]
\centering \scriptsize \caption{Same as Table.~\ref{Table 2}, but
for $^{24}$Ne, $^{24}$Na and $^{24}$Mg.}\label{Table 6}
\resizebox{\columnwidth}{!}{%
\begin{tabular}{|c|c|cc|cc|cc|}
\hline $\rho$$\it Y_{e}$ & T (K) & \multicolumn{2}{c|}{$^{24}$Ne}&
\multicolumn{2}{c|}{$^{24}$Na} & \multicolumn{2}{c|}{$^{24}$Mg}\\
\cline{3-8}(g.cm$^{-3}$) & &$\lambda_{\bar\nu}$ & $\lambda_{\nu}$ &
$\lambda_{\bar\nu}$ & $\lambda_{\nu}$ &$\lambda_{\bar\nu}$ & $\lambda_{\nu}$\\
\hline
10$^{1}$& 1.00$\times10^{7}$& 3.68$\times10^{-3}$& 1.00$\times10^{-100 }$& 4.69$\times10^{-6}$& 1.00$\times10^{-100 }$& 1.00$\times10^{-100 }$& 1.00$\times10^{-100 }$\\
 10$^{1}$& 4.00$\times10^{8}$& 3.68$\times10^{-3}$& 1.00$\times10^{-100 }$& 4.69$\times10^{-6}$& 1.18$\times10^{-52 }$& 1.00$\times10^{-100 }$& 6.67$\times10^{-84 }$\\
 10$^{1}$& 7.00$\times10^{8}$& 3.68$\times10^{-3}$& 1.00$\times10^{-100 }$& 4.78$\times10^{-6}$& 2.29$\times10^{-33 }$& 1.00$\times10^{-100 }$& 1.28$\times10^{-51 }$\\
 10$^{1}$& 1.00$\times10^{9}$& 3.68$\times10^{-3}$& 8.28$\times10^{-76 }$& 5.46$\times10^{-6}$& 1.90$\times10^{-24 }$& 6.21$\times10^{-75 }$& 1.83$\times10^{-37 }$\\
 10$^{1}$& 2.00$\times10^{9}$& 3.78$\times10^{-3}$& 4.02$\times10^{-39 }$& 5.11$\times10^{-5}$& 2.06$\times10^{-13 }$& 5.56$\times10^{-38 }$& 2.94$\times10^{-20 }$\\
 10$^{1}$& 3.00$\times10^{9}$& 4.67$\times10^{-3}$& 1.57$\times10^{-26 }$& 5.22$\times10^{-4}$& 2.33$\times10^{-9}$& 1.88$\times10^{-25 }$& 4.43$\times10^{-14 }$\\
\hline
\end{tabular}%
}
\end{table*}
%

\begin{table*}[htb]
\centering \tiny \caption{Calculated $\gamma$-ray heating and
antineutrino (neutrino) cooling rates due to weak rates on $^{25}$Na
and $^{25}$Mg at different selected densities and temperatures in
stellar environment. The first column shows the stellar densities
($\rho$Y$_{e}$) (in units of g.cm$^{-3}$). The rates are tabulated
in units of MeV.s$^{-1}$.}\label{Table 7}\resizebox{\columnwidth}{!}{%
\begin{tabular}{|c|c|cccc|cccc|}
\hline $\rho$$\it Y_{e}$ & T (K)  &
\multicolumn{4}{c|}{$^{25}$Na}&\multicolumn{4}{c|}{$^{25}$Mg}\\
\cline{3-10} (g.cm$^{-3}$)& &$\lambda_{\gamma}^{\beta}$ &
$\lambda_{\gamma}^{EC}$
& $\lambda_{\bar\nu}$ & $\lambda_{\nu}$ &$\lambda_{\gamma}^{\beta}$ & $\lambda_{\gamma}^{EC}$&$\lambda_{\bar\nu}$ & $\lambda_{\nu}$\\
\hline
10$^{1}$& 1.00$\times10^{7}$& 1.51$\times10^{-2}$& 1.00$\times10^{-100 }$& 3.46$\times10^{-2}$& 1.00$\times10^{-100 }$& 1.00$\times10^{-100 }$& 1.00$\times10^{-100 }$& 1.00$\times10^{-100 }$& 1.00$\times10^{-100 }$\\
 10$^{1}$& 4.00$\times10^{8}$& 1.51$\times10^{-2}$& 1.00$\times10^{-100 }$& 3.46$\times10^{-2}$& 1.00$\times10^{-100 }$& 2.21$\times10^{-73 }$& 5.35$\times10^{-68 }$& 5.31$\times10^{-64 }$& 5.62$\times10^{-59 }$\\
 10$^{1}$& 7.00$\times10^{8}$& 1.51$\times10^{-2}$& 4.72$\times10^{-62 }$& 3.47$\times10^{-2}$& 2.43$\times10^{-64 }$& 6.00$\times10^{-44 }$& 1.44$\times10^{-43 }$& 2.90$\times10^{-38 }$& 4.85$\times10^{-37 }$\\
 10$^{1}$& 1.00$\times10^{9}$& 1.52$\times10^{-2}$& 1.33$\times10^{-44 }$& 3.52$\times10^{-2}$& 2.52$\times10^{-46 }$& 4.04$\times10^{-32 }$& 1.71$\times10^{-31 }$& 9.59$\times10^{-28 }$& 4.59$\times10^{-27 }$\\
 10$^{1}$& 2.00$\times10^{9}$& 1.59$\times10^{-2}$& 1.17$\times10^{-23 }$& 4.26$\times10^{-2}$& 7.82$\times10^{-25 }$& 3.05$\times10^{-17 }$& 6.65$\times10^{-17 }$& 3.29$\times10^{-15 }$& 7.85$\times10^{-15 }$\\
 10$^{1}$& 3.00$\times10^{9}$& 1.70$\times10^{-2}$& 2.45$\times10^{-16 }$& 5.36$\times10^{-2}$& 2.84$\times10^{-17 }$& 5.26$\times10^{-12 }$& 9.29$\times10^{-12 }$& 9.42$\times10^{-11 }$& 2.18$\times10^{-10 }$\\
\hline
\end{tabular}%
}
\end{table*}

\begin{table*}[ht!]
\centering \scriptsize \caption{Comparison of pn-QRPA calculated
$\gamma$-ray heating rates and neutrino cooling rates (this work)
with previous calculations due to electron capture on $^{23}$Na, as
function of stellar density and temperature. ADNDT99 implies
\citep{Nabi99}, OHMTS signifies \citep{Oda94} calculation whereas
STN15 represent \citep{Suz16} results.} \label{Table 8}
\resizebox{\columnwidth}{!}{%
\begin{tabular}{|c|c|cccc|cccc|}
\hline $\rho$$\it Y_{e}$ &T(K) &
\multicolumn{4}{c|}{$\gamma$-ray heating rates (MeV.s$^{-1}$)} & 
\multicolumn{4}{c|}{Neutrino cooling rates (MeV.s$^{-1}$)}\\
\cline{3-10} (g.cm$^{-3}$)&  &This Work &ADNDT99   &OHMTS     &STN15     &This Work  &ADNDT99    &OHMTS     &STN15\\
\hline
       & $1.00\times10^{7} $ & $<10^{-100}$        &$<10^{-100}$   & $<10^{-100}$     & $<10^{-100}$     & $<10^{-100}$&$<10^{-100}$     & $<10^{-100}$     & $<10^{-100}$ \\
       & $1.00\times10^{8} $ & $<10^{-100}$        &$<10^{-100}$    & $<10^{-100}$     & $<10^{-100}$     & $<10^{-100}$&$<10^{-100}$     & $<10^{-100}$     & $<10^{-100}$ \\
$10^7$ & $1.00\times10^{9} $ & $4.17\times10^{-25}$&$9.77\times10^{-29}$   &$5.12\times10^{-29}$& $2.88\times10^{-29}$&  $8.93\times10^{-24}$&$2.70\times10^{-24}$ &$8.41\times10^{-25}$& $6.07\times10^{-25}$\\
       & $1.00\times10^{10}$ & $1.12\times10^{-3}$ &$7.78\times10^{-4}$ &$6.93\times10^{-4}$&--& $5.29\times10^{-3}$&$3.36\times10^{-3}$ &$1.89\times10^{-3}$& -- \\
       & $3.00\times10^{10}$ & $2.32\times10^{1}$  & $6.28\times10^{0}$  &  $8.99\times10^{1}$   & --       & $1.49\times10^{2}$&$8.28\times10^{1}$&   $1.65\times10^{2}$  & --  \\
\hline
       & $1.00\times10^{7}$  & $<10^{-100}$& $<10^{-100}$   & $<10^{-100}$     & $<10^{-100}$     & $<10^{-100}$ &$<10^{-100}$    & $<10^{-100}$     & $<10^{-100}$ \\
\hline
 \end{tabular}%
 }
\end{table*}

\begin{table*}[ht!]
\centering \scriptsize \caption{Same as Table.~\ref{Table 8} but for
$^{23}$Ne.} \label{Table 9}
\resizebox{\columnwidth}{!}{%
\begin{tabular}{|c|c|cccc|cccc|}
\hline $\rho$$\it Y_{e}$&T(K) & \multicolumn{4}{c|}{$\gamma$-ray
heating rates (MeV.s$^{-1}$)} \vline & \multicolumn{4}{c|}{Neutrino
cooling rates (MeV.s$^{-1}$)}\\
\cline{3-10}(g.cm$^{-3}$)&  &This Work &ADNDT99   &OHMTS     &STN16     &This Work &ADNDT99    &OHMTS     &STN16\\
\hline
      & $1.00\times10^{7}$  & $<10^{-100}$&$<10^{-100}$& $<10^{-100}$& $<10^{-100}$& $<10^{-100}$&$<10^{-100}$& $<10^{-100}$& $<10^{-100}$ \\
       & $1.00\times10^{8}$ & $<10^{-100}$&$<10^{-100}$& $<10^{-100}$& $<10^{-100}$& $<10^{-100}$&$<10^{-100}$& $<10^{-100}$& $<10^{-100}$ \\
$10^7$& $1.00\times10^{9}$& 4.79$\times10^{-45}$&2.33$\times10^{-45}$ &3.40$\times10^{-53}$& 3.71$\times10^{-55}$& 1.26$\times10^{-43}$&4.73$\times10^{-44}$ & 1.10$\times10^{-44}$& 1.18$\times10^{-45}$\\
       & $1.00\times10^{10}$ & 5.19$\times10^{-05}$&6.24$\times10^{-06}$ &7.60$\times10^{-06}$& --   & 2.93$\times10^{-04}$&8.22$\times10^{-05}$ &3.55$\times10^{-05}$& -- \\
       & $3.00\times10^{10}$& 1.18$\times10^{01}$&3.26$\times10^{00}$ &1.30$\times10^{01}$& --   & 8.49$\times10^{01}$&1.71$\times10^{01}$ & 2.36$\times10^{01}$ & --  \\
\hline
      & $1.00\times10^{7}$  & $<10^{-100}$& $<10^{-100}$ & $<10^{-100}$& $<10^{-100}$& $<10^{-100}$&$<10^{-100}$& $<10^{-100}$ & $<10^{-100}$ \\
\hline
 \end{tabular}%
 }
\end{table*}
\begin{table*}[ht!]
\centering \scriptsize \caption{Same as Table.~\ref{Table 8} but for
$^{25}$Mg.} \label{Table 10} 
\resizebox{\columnwidth}{!}{%
\begin{tabular}{|c|c|cccc|cccc|}
\hline $\rho$$\it Y_{e}$&T(K) & \multicolumn{4}{c|}{$\gamma$-ray
heating rates (MeV.s$^{-1}$)} \vline & \multicolumn{4}{c|}{Neutrino
cooling
rates (MeV.s$^{-1}$)}\\
\cline{3-10}(g.cm$^{-3}$) & &This Work & ADNDT99 & OHMTS & STN16 & This Work & ADNDT99 & OHMTS & STN16\\
\hline
 & {1.00$\times10^{7}$}  & {$<10^{-100}$} & {$<10^{-100}$} & {$<10^{-100}$} & {$<10^{-100}$} & {$<10^{-100}$} & {$<10^{-100}$} & {$<10^{-100}$} & {$<10^{-100}$} \\
 & {1.00$\times10^{8}$} & {$<10^{-100}$} & {$<10^{-100}$} & {$<10^{-100}$} & {$<10^{-100}$} & {$<10^{-100}$} & {$<10^{-100}$} & {$<10^{-100}$} & {$<10^{-100}$} \\
$10^{7}$ & {1.00$\times10^{9}$} & {1.90$\times10^{-25}$}&{1.44$\times10^{-25}$} & {3.06$\times10^{-23}$} & {2.08$\times10^{-23}$} & {5.13$\times10^{-21}$}&{4.52$\times10^{-22}$} & {5.30$\times10^{-22}$} & {3.50$\times10^{-22}$} \\
 & {1.00$\times10^{10}$} & {3.40$\times10^{-3}$}&{1.94$\times10^{-3}$} & {9.89$\times10^{-4}$} & -- & {1.46$\times10^{-2}$}&{3.92$\times10^{-3}$} & {3.10$\times10^{-3}$} & -- \\
 & {3.00$\times10^{10}$} & {5.81$\times10^{1}$}&{3.54$\times10^{1}$} & {9.29$\times10^{1}$} & -- & {1.88$\times10^{2}$}&{1.08$\times10^{2}$} & {1.69$\times10^{2}$} & -- \\
\hline
 & {1.00$\times10^{7}$}  & {$<10^{-100}$} & {$<10^{-100}$} & {$<10^{-100}$} & {$<10^{-100}$} & {$<10^{-100}$} & {$<10^{-100}$} & {$<10^{-100}$} & {$<10^{-100}$} \\
\hline
\end{tabular}%
}
\end{table*}

\begin{table*}[ht!]
\centering \scriptsize \caption{Same as Table.~\ref{Table 8} but for
$^{25}$Na.} \label{Table 11}
\resizebox{\columnwidth}{!}{%
\begin{tabular}{|c|c|cccc|cccc|}

 \hline $\rho$$\it Y_{e}$&T(K) &
\multicolumn{4}{c|}{$\gamma$-ray heating rates (MeV.s$^{-1}$)} & 
\multicolumn{4}{c|}{Neutrino cooling
rates (MeV.s$^{-1}$)}\\
\cline{3-10}(g.cm$^{-3}$) & &This Work &ADNDT99 &OHMTS &STN16 &This Work &ADNDT99 &OHMTS &STN16\\
\hline
& {1.00$\times10^{7}$}  & {$<10^{-100}$} & {$<10^{-100}$} & {$<10^{-100}$} & {$<10^{-100}$} & {$<10^{-100}$} & {$<10^{-100}$} & {$<10^{-100}$} & {$<10^{-100}$}\\
 & {1.00$\times10^{8}$} & {$<10^{-100}$} & {$<10^{-100}$} & {$<10^{-100}$} & {$<10^{-100}$} & {$<10^{-100}$} & {$<10^{-100}$} & {$<10^{-100}$} & {$<10^{-100}$} \\
10$^{7}$ & {1.00$\times10^{9}$} & {2.81$\times10^{-40}$}&{5.07$\times10^{-39}$} & {3.98$\times10^{-46}$} & {5.01$\times10^{-47}$} & {1.49$\times10^{-38}$}&{3.49$\times10^{-38}$} & {3.98$\times10^{-39}$} & {2.51$\times10^{-39}$} \\
 & {1.00$\times10^{10}$} & {2.42$\times10^{-4}$}&{9.18$\times10^{-5}$} & {3.48$\times10^{-5}$} & -- & {4.42$\times10^{-4}$}&{3.62$\times10^{-4}$} & {7.23$\times10^{-5}$} & -- \\
 & {3.00$\times10^{10}$} & {3.03$\times10^{1}$}&{1.10$\times10^{1}$} & {1.79$\times10^{1}$} & -- & {1.46$\times10^{2}$}&{1.07$\times10^{2}$} & {3.53$\times10^{1}$} & -- \\
\hline
& {1.00$\times10^{7}$}  & {$<10^{-100}$} & {$<10^{-100}$} & {$<10^{-100}$} & {$<10^{-100}$} & {$<10^{-100}$} & {$<10^{-100}$} & {$<10^{-100}$} & {$<10^{-100}$} \\
\hline
\end{tabular}%
}
\end{table*}
\begin{table*}[!htbp]
\centering \scriptsize \caption{Comparison of pn-QRPA calculated
$\gamma$-ray heating rates and neutrino cooling rates (this work)
with previous calculations, due to $\beta^{-}$-decay from $^{20}$O,
as function of stellar density and temperature. OHMTS signifies
\citep{Oda94} calculation whereas STN15 represent \citep{Suz16}
results.}\label{Table 12}
\begin{tabular}{|c|c|ccc|ccc|}
\hline $\rho$$\it Y_{e}$&T(K) &
\multicolumn{3}{c|}{$\gamma$-ray heating rates (MeV.s$^{-1}$)} & 
\multicolumn{3}{c|}{Neutrino cooling
rates (MeV.s$^{-1}$)}\\
\cline{3-8}(g.cm$^{-3}$) & &This Work  &OHMTS &STN16 &This Work &OHMTS &STN16\\
\hline
& {1.00$\times10^{7}$}  & {6.12$\times10^{-2}$} & {5.26$\times10^{-2}$} & -- & {$<10^{-100}$} & {7.74$\times10^{-2}$} & --\\
 & {1.00$\times10^{8}$} & {6.12$\times10^{-2}$} & {5.26$\times10^{-2}$} & -- & {$<10^{-100}$} & {7.74$\times10^{-2}$} & -- \\
10$^{1}$ & {1.00$\times10^{9}$} & {6.12$\times10^{-2}$}&{5.26$\times10^{-2}$} & -- & {1.58$\times10^{-26}$} & {7.74$\times10^{-2}$}& -- \\
 & {1.00$\times10^{10}$} & {1.46$\times10^{0}$}&{1.10$\times10^{0}$} & -- & {6.37$\times10^{-3}$}&{1.58$\times10^{0}$}& -- \\
 & {3.00$\times10^{10}$} & {2.16$\times10^{2}$}&{7.92$\times10^{2}$} & -- & {1.06$\times10^{2}$}&{1.04$\times10^{3}$}& -- \\
\hline
& {1.00$\times10^{7}$}  & {6.12$\times10^{-2}$} & {5.26$\times10^{-2}$} & -- & {$<10^{-100}$} & {7.74$\times10^{-2}$} & -- \\
\hline
\end{tabular}%
\end{table*}

\begin{thebibliography}{99}

\bibitem[Wang et al. (2012)]{Wang12} M. Wang, G. Audi, A.H. Wapstra, F.G. Kondev, M. MacCormick, X. Xu, and B. Pfeiffer B, Chin. Phys. C \textbf{36}  (2012) 1603.
\bibitem[M\"{o}ller et al. (2016)]{Mol16}P. M\"{o}ller A. J. Sierk, T. Ichikawa, and H. Sagawa, At. Data Nucl. Data Tables ${\bf109}$ (2016) 1-204.
\bibitem[Raman et al. (2001)]{Ram01}S. Raman, C. W. Nestor, and P. Tikkanen. At. Data Nucl. Data Tables ${\bf78 (1)}$ (2001) 1-128.
\bibitem[Woosley et al. (2002)]{Woosley02}S. E. Woosley, A. Heger, and T. A. Weaver,  Rev. Mod. Phys. ${\bf74}$ (2002) 1015.
\bibitem[Halabi et al. (2014)]{Halabi14}M. Halabi, Ghina, and Mounib El Eid. AIP Conf. Proc. ${\bf1645 (1)}$ (2015) 339-343.
\bibitem[Ryan (2010)]{Ryan2010}S. G. Ryan, A. J. Norton,  Stellar Evolution and Nucleosynthesis. Cambridge University Press. (2010) 135-136. ISBN 978-0-521-13320-3.
\bibitem[Miyaji et al. (1980)]{Miyaji80} S. Miyaji, K. Nomoto,K. Yokoi, and D. Sugimoto,  Publ. Astron. Soc. Japan ${\bf32}$, (1980) 303.
\bibitem[Woosley \& Weaver (1980)]{W&W80} S. E. Woosley, and T. A. Weaver,  Astrophys. J. ${\bf238}$ (1980) 1017-1025.
\bibitem[Nomoto(1987)]{Nomoto87} K. Nomoto   Astrophys. J. ${\bf322}$ (1987) 206-214.
\bibitem[Guti\'{e}rrez et al. (2005)]{Gut05} J. Guti\'{e}rrez, R. Canal, and E. Garc\'{\i}a-Berro. Astron. \& Astrophys. ${\bf435 (1)}$ (2005) 231-237.
\bibitem[Miyaji \& Nomoto (1987)]{M&N87} S. Miyaji, and K. Nomoto, Astrophys. J. ${\bf318}$ (1987) 307-315.
\bibitem[Takahara et al. (1989)]{Takahara89}M. Takahara, M. Hino, T. Oda, K. Muto, A. A. Wolters, P. W. M. Glaudemans, and K. Sato, Nucl. Phys. A ${\bf504 (1)}$ (1989) 167-192.
\bibitem[Wildenthal (1984)]{Wil84} B. H. Wildenthal. Prog. Part. Nucl. Phy., edited by D. H. Wilkinson (Pergamon, Oxford) ${\bf11}$ (1984) 5.
\bibitem[Brown \& Richter (2006)]{Bro06} B. A. Brown and W. A. Richter, Phys. Rev. C ${\bf 74}$ (2006) 034315.
\bibitem[Richter et al. (2008)]{Ric08} W. A. Richter, S. Mkhize, and B. A. Brown, Phys. Rev. C ${\bf 78}$ (2008) 064302.
\bibitem[Garc\'{i}a et al. (1997)]{Gar97} E. Garc\'{i}a-Berro, C. Ritossa and I. Iben,  Astrophys. J. ${\bf485 (2)}$ (1997) 765
\bibitem[Gupta et al. (2007)]{Gupta07} S. Gupta, E. F. Brown, H. Schatz, P. M$\ddot{o}$ller, and K. -L. Kratz, Astrophys. J. ${\bf662 (2)}$ (2007) 1188.
\bibitem[Jones et al. (2013)]{Jones13} S. Jones, R. Hirschi, K.I. Nomoto, T. Fischer, F.X. Timmes, F. Herwig, B. Paxton, H. Toki, T. Suzuki, G. Martinez-Pinedo,  and
Y.H.Lam, Astrophys. J. ${\bf772 (2)}$ (2013) 150.
\bibitem[Mart\'{i}nez-Pinedo et al. (2014)]{Mar14} G. Mart\'{i}nez-Pinedo, Y. H. Lam, K. Langanke, R. G. T. Zegers, and C. Sullivan, Phys. Rev. C ${\bf89 (4)}$ (2014) 045806.
\bibitem[Suzuki et al. (2016)]{Suz16} T. Suzuki, H. Toki, and K. I. Nomoto, The Astrophy. J. ${\bf817 (2)}$ (2016) 163.
\bibitem[Oda et al. (1994)]{Oda94} T. Oda, M. Hino, K. Muto, M. Takahara, and K. Sato, At. Data Nucl. Data Tables ${\bf56}$ (1994) 231-403.
\bibitem[Nabi \&  Klapdor-Kleingrothaus (1999)]{Nabi99} J.-U. Nabi and H.V. Klapdor-Kleingrothaus, At. Data Nucl. Data Tables ${\bf 71}$ (1999) 149.
\bibitem[Nabi \&  Klapdor-Kleingrothaus  (1999a)]{Nabi99a} J.-U. Nabi and H.V. Klapdor-Kleingrothaus, Eur. Phys. J. A ${\bf5}$ (1999) 337
\bibitem[Nabi\&  Klapdor-Kleingrothaus  (2004)]{Nabi04} J.-U. Nabi and H.V. Klapdor-Kleingrothaus, At. Data Nucl. Data Tables ${\bf88}$ (2004) 237
\bibitem[J. -U. Nabi (2008a)]{Nabi08a} J.-U. Nabi, Physica Scripta ${\bf78 (3)}$ (2008) 035201.
\bibitem[J. -U. Nabi (2008)]{Nabi08} J.-U. Nabi, Phys. Rev. C ${\bf78 (4)}$ (2008) 045801.
\bibitem[Muto et al. (1992)]{mut92} K. Muto, E. Bender, T. Oda and H. V. Klapdor, Z. Phys. A ${\bf 341}$ (1992)
407.
\bibitem[Nabi \& Rahman (2007)]{Nabi07} J.-U. Nabi and M.-U. Rahman, Phys. Rev. C ${\bf75 (3)}$ (2007) 035803.
\bibitem[Nabi \& Boyukata (2017)]{Nabi17}J. -U. Nabi and M. B\"{o}y\"{u}kata Astrophys. Space Sci. ${\bf362 (1)}$ (2017) 9.
\bibitem[Staudt et al. (1990)] {Sta90} A. Staudt, E. Bender, K. Muto and H.V. Klapdor-Kleingrothaus, At. Data Nucl. Data Tables ${\bf 44}$ (1990) 79.
\bibitem[Nakamura et al. (2010)]{Nak10} K. Nakamura,  and Particle Data Group, J. Phys. G: Nucl. and Part. Phys. ${\bf37 (7A)}$ (2010) 075021.
\bibitem[Rodin et al. (2006)]{Rod06} V. Rodin, A. Faessler, F. Simkovic, and P. Vogel, Czech. J. Phys. \textbf{56} (2006) 495.
\bibitem[Fuller et al. (1980)]{Ful80} G. M. Fuller, W. A. Fowler, and M. J. Newman, Astrophys. J. Supp. Ser. \textbf{42} (1980) 447.
\bibitem[Fuller et al. (1982)]{Ful82} G. M. Fuller, W. A. Fowler, and M. J. Newman, Astrophys. J. \textbf{252} (1982) 715.
\bibitem[Fuller et al. (1982a)]{Ful82a} G. M. Fuller, W. A. Fowler, and M. J. Newman, Astrophys. J. Supp. Ser. \textbf{48} (1982) 279.
\bibitem[Fuller et al. (1985)]{Ful85} G. M. Fuller, W. A. Fowler, and M. J. Newman, Astrophys. J. \textbf{293} (1985) 1.
\bibitem[Hardy et al. (2009)]{Har09} J. C. Hardy and I. S. Towner, Phys. Rev. C ${\bf79 (5)}$ (2009) 055502.
\bibitem[Gove and Martin(1971)]{Gov71} N. B. Gove and M. J. Martin, At. Data Nucl. Data Tables, \textbf{10} (1971) 205.
\bibitem[Nilsson (1955)]{Nil55} S. G. Nilsson, Mat. Fys. Medd. Dan. Vid. Selsk \textbf{ 29}  (1955) 16.
\bibitem[Hirsch et al. (1991)]{Hir91} M. Hirsch, A. Staudt, K. Muto, and H. V. Klapdor-Kliengrothaus, Nucl. Phys. A \textbf{535} (1991) 62.
\bibitem[Ikeda (1964)]{Ike64} I. Ikeda, Prog. Theor. Phys., \textbf{31} (1964)
434.
\end{thebibliography}
\end{document}